\documentclass[prresearch,twocolumn,aps,longbibliography]{revtex4-2}

\usepackage{amsmath}
\usepackage{amsfonts}
\usepackage{amssymb}
\usepackage[usenames,dvipsnames]{color}
\usepackage[colorlinks=true,linkcolor=Maroon,citecolor=Maroon,urlcolor=Maroon]{hyperref}
\usepackage{graphicx}
\usepackage{xcolor}
\usepackage{comment}

\usepackage{enumitem}
\usepackage{amsthm}          
\usepackage{tikz,pgfplots}

\newcommand{\E}{\mathbb{E}}

\makeatletter
\def\ps@pprintTitle{%
 \let\@oddhead\@empty 
 \let\@evenhead\@empty
 \def\@oddfoot{}%
 \let\@evenfoot\@oddfoot}
 \makeatother 

\newcommand{\A}{\mathbf{A}}
\newcommand{\B}{\mathbf{B}}
\newcommand{\R}{\mathbf{R}}
\newcommand{\K}{\mathbf{K}}

\newcommand{\PP}{\mathbf{P}}

\newcommand{\I}{\mathbf{I}}
\newcommand{\Q}{\mathbf{Q}}
\newcommand{\U}{\mathbf{U}}
\newcommand{\V}{\mathbf{V}}
\newcommand{\D}{\mathbf{D}}
\newcommand{\EE}{\mathbf{E}}

\newcommand{\x}{\mathbf{x}}
\newcommand{\rr}{\mathbf{r}}
\newcommand{\el}{\boldsymbol{\ell}}
\newcommand{\w}{\mathbf{w}}
\newcommand{\e}{\mathbf{e}}
\newcommand{\f}{\mathbf{f}}

\newcommand{\vv}{\mathbf{\hat{v}}}
\newcommand{\uu}{\mathbf{\hat{u}}}

\newcommand{\et}{\boldsymbol{\eta}}
\newcommand{\PPi}{\boldsymbol{\Pi}}
\newcommand{\Sig}{\boldsymbol{\Sigma}}
\newcommand{\Lamb}{\boldsymbol{\Lambda}}

\begin{document}

    \title{Non-Normal Eigenvector Amplification in Multi-Dimensional Kesten Processes
            }

    \author{Virgile Troude}
    \author{Didier Sornette}
    \affiliation{
        Institute of Risk Analysis, Prediction and Management (Risks-X),
        Academy for Advanced Interdisciplinary Sciences,
        Southern University of Science and Technology, Shenzhen, China
    }

   \begin{abstract}
Heavy-tailed fluctuations and power law statistics pervade physics, finance, and economics, yet their origin is often ascribed to systems poised near criticality.  
Here we show that such behavior can emerge far from instability through a universal mechanism of \emph{non-normal eigenvector amplification} in multidimensional Kesten processes $\x_{t+1}=\A_t\x_t+\et_t$,
where $\A_t$ are random interaction matrices and $\et_t$ represents external inputs, capturing the evolving interdependence among $N$ coupled components.  
Even when each random multiplicative matrix is spectrally stable, non-orthogonal eigenvectors generate transient growth that renormalizes the Lyapunov exponent and lowers the tail exponent, producing stationary power laws without eigenvalues crossing the stability boundary.  
We derive explicit relations linking the Lyapunov exponent and the tail index to the statistics of the condition number, $\gamma\!\sim\!\gamma_0+\ln\kappa$ and $\alpha\!\sim\!-2\gamma/\sigma_\kappa^2$, confirmed by numerical simulations.  
This framework offers a unifying geometric perspective that help interpret diverse phenomena, including polymer stretching in turbulence, magnetic field amplification in dynamos, volatility clustering and wealth inequality in financial systems.  
Non-normal interactions provide a collective route to scale-free behavior in globally stable systems, defining a new universality class where multiplicative feedback and transient amplification generate critical-like statistics without spectral criticality.
\end{abstract}

    \maketitle


    \section{Introduction}


    Heavy-tailed statistics pervade complex systems, appearing in phenomena as diverse as turbulent flows, financial markets, ecological dynamics, and biological growth.  
    A widely used framework for this mechanism is the class of Kesten-type stochastic recursions~\cite{kesten1973random}, 
    where multiplicative growth combined with additive noise yields stationary power-law tails.
    These \emph{Kesten processes} form a unifying framework for scale-free behavior across disciplines, 
    from disordered transport and self-organized criticality~\cite{deCalan1985,Takayasu1997,BlankSolomon2000} 
    to population dynamics~\cite{Statman2014}, 
    macroeconomic fluctuations~\cite{LuxSornette2002,lux2006financial,Gabaix2009,Derman2010}, 
    and wealth inequality~\cite{Sornette1998,benhabib2016wealth}.

    Traditionally, the origin of power laws in such models is traced to episodes of \emph{spectral supercriticality}, 
    when eigenvalues of the random multiplicative operator temporarily exceed the stability boundary.  
    Heavy tails are then viewed as the signature of rare supercritical bursts moderated by global mean reversion.  
    Yet, this explanation overlooks a distinct geometric aspect of linear dynamics: 
    even when all eigenvalues indicate stability, non-orthogonal eigenvectors can transiently amplify fluctuations.  

    This property, known as \emph{non-normality}, has long been recognized in hydrodynamic stability theory, 
    where it accounts for transient energy growth in linearly stable shear flows~\cite{Trefethen1993,FarrellIoannou1996a,chomaz2005global,schmid2007nonmodal,reddy1993energy}.  
    Non-normal operators can exhibit responses several orders of magnitude stronger than predicted by eigenvalue analysis alone~\cite{Embree2005}, 
    and stochastic forcing can sustain persistent variability through such amplification mechanisms~\cite{FarrellIoannou1996b,FarrellIoannou2003}.  
    More broadly, non-normal amplification has emerged as a generic organizing principle of complex systems~\cite{troude2025}, 
    governing phenomena ranging from turbulence~\cite{chomaz2005global,schmid2007nonmodal} to neural and ecological networks~\cite{allesina2015stability,asllani2018structure}.  

    In the companion Letter~\cite{troude2025letter}, we introduced a new and general mechanism linking non-normality to the statistics of multiplicative stochastic processes.  
    There, we showed that even in the absence of spectral instability,
    \emph{non-normal eigenvector amplification} generically produces power-law stationary distributions.  
    When random matrices are non-normal (not unitarily diagonalizable), transient growth occurs through the coupling of nearly aligned eigenvectors,
    allowing momentary bursts of amplification even when all eigenvalues lie strictly within the unit circle.  
    This effect provides a distinct and generic route to heavy-tailed statistics, independent of classical spectral criticality.

    The present paper provides the detailed theoretical derivations, numerical verifications, and empirical analyses supporting the results of the Letter~\cite{troude2025letter}.  
    We develop a quantitative framework connecting the Lyapunov exponent $\gamma$ and the tail exponent $\alpha$ to the statistical properties of eigenvector geometry,
    characterized by the matrix condition number $\kappa$.  
    Specifically, we show that non-normality leads to a renormalization of the effective Lyapunov exponent,
    $\gamma \simeq \gamma_0 + \E\left[\ln \kappa \right]$, and a corresponding modification of the tail exponent,
    $\alpha \sim -2\gamma / \sigma_\kappa^2$, where $\sigma_\kappa^2$ denotes the variance of $\ln \kappa$.  
    These relations demonstrate how transient amplification due to eigenvector non-orthogonality directly shapes stationary heavy-tailed statistics.

    The implications of this finding are broad.
    Non-normal amplification provides a new form of \emph{universality} in the emergence of power laws:
    fat tails may arise not only from rare excursions into instability but also from generic geometrical properties of high-dimensional random operators.  
    We illustrate this mechanism across multiple domains.
    In turbulent polymer solutions, it elucidates how intermittent stretching events emerge from transient alignments of the velocity-gradient eigenvectors.  
    In magnetohydrodynamics, it clarifies the kinematic stage of the small-scale turbulent dynamo,
    where magnetic-field amplification follows the same geometric principle.  
    In financial systems, it provides a natural multidimensional generalization of Kesten-GARCH-type dynamics that rationalizes the near-universal inverse-cubic law
    ($\alpha \!\approx\! 3$) observed across assets and markets without invoking exogenous heavy-tailed shocks.

    By revealing non-normal eigenvector amplification as a fundamental and universal mechanism for the emergence of heavy tails,
    our work unifies previously disconnected phenomena, from fluid instabilities and dynamo action to financial volatility clustering, under a single theoretical framework.  
    It thereby extends the classical theory of Kesten processes and situates non-normality as a central organizing principle in the dynamics and statistics of complex systems.

The paper is organized as follows. Section 2 introduces the general mathematical framework. Section 3 examines the two-dimensional case, which provides analytical tractability and offers key insights into the underlying mechanism. Section 4 extends the quantitative analysis to the general 
$N$-dimensional setting. Section 5 illustrates the relevance of the results through two physical examples (polymer stretching in turbulent flows and the small-scale turbulent dynamo) and four applications in finance and economics (volatility models of the ARCH type, cointegration models, factor models, and wealth inequality dynamics). Section 6 summarizes the main findings and outlines perspectives for future work.


    \section{Mathematical Framework}


    We begin by formalizing the multidimensional Kesten process and recalling classical results. 
    This section provides the necessary preliminaries before turning to the role of non-normality in driving critical behavior.

    \subsection{Multidimensional Kesten Process}

    We consider a generalized $N$-dimensional Kesten process $(\x_t)_{t\geq 0}$ defined by
    \begin{equation}    \label{eq:apx_kesten}
        \x_{t+1} = \A_t \x_t + \sqrt{2\delta}\,\et,
        \qquad \et \overset{\textit{i.i.d.}}{\sim} \mathcal{N}(0,\I),
    \end{equation}
    where the random matrices $(\A_t)$ are assumed to be i.i.d.

    Following the framework introduced by Furstenberg and Kesten~\cite{FurstenbergKesten1960},  
    we define the product norm
    \begin{equation}    \label{eq:apx_matrix_prod}
        \pi_t = \|\PPi_t\|,
        \qquad 
        \PPi_t = \prod_{s=1}^t \A_s,
    \end{equation}
    where $\|\cdot\|$ denotes the matrix norm, taken here to be the $L_2$-norm.  
    The associated Lyapunov exponent is then
    \begin{equation}    \label{eq:apx_lyapunov}
        \gamma := \lim_{t\to\infty} \frac{1}{t}\,
        \mathbb{E}\!\left[\ln \pi_t\right].
    \end{equation}

    A classical sufficient condition for the existence of a unique stationary solution  
    (and for the recursion to be geometrically ergodic) is the ``average contraction'' condition $\gamma < 0$,  
    together with the non-degeneracy condition $\mathbb{P}[\et = 0] < 1$.  
    In this case, the stationary forward solution is given by
    \begin{equation}
        \x \;\overset{\text{d}}{=}\; 
        \lim_{T\to\infty} \sqrt{2\delta}
        \sum_{t=0}^T \PPi_t \et_{T-t}.
    \end{equation}
    The intuition is that when $\gamma < 0$, the products $\PPi_t$ decay exponentially fast to zero.  
    Conversely, if $\gamma > 0$, the process diverges exponentially and no stationary distribution exists.

    A second fundamental result concerns the tail behavior of the stationary solution.  
    Under the following assumptions:
    \begin{itemize}
        \item Global stability: $\gamma < 0$;
        \item Local instability: $\mathbb{P}\!\left[\|\A_t\| > 1\right] > 0$;
        \item A mild non-arithmetic (non-lattice) condition on the distribution of $\ln\|\A_t\|$;
        \item  Defining the moment generating function of the random matrix product
        $\Pi_t = \prod_{s=1}^{t}A_s$ through
        \begin{equation}
            \begin{split}
            \phi(\alpha)
            &:= \lim_{t\to\infty}
            \frac{1}{t}\,
            \ln
            \mathbb{E}\!\left[
                \big\|\Pi_t\big\|^{\alpha}
            \right] \\
            &= 
            \lim_{t\to\infty}
            \frac{1}{t}\,
            \ln
            \mathbb{E}\!\left[
                \exp\!\Big(
                    \alpha
                    \ln\big\|\textstyle\prod_{s=1}^{t}A_s\big\|
                \Big)
            \right].
            \end{split}
            \label{eq:phi_defrtbt}
        \end{equation}
        there exists an $\alpha > 0$ such that $ \phi(\alpha)=0$ and $\mathbb{E}\!\left[\|\et_t\|^\alpha\right] < \infty$;
            \end{itemize}
        then, the stationary distribution exhibits a power law tail of the form
            \begin{equation}
                \mathbb{P}\!\left[\,\|\x\| > x\,\right] \sim C x^{-\alpha},
                \qquad x \to \infty.
                \label{dfnhtyq}
            \end{equation}

    Moreover, the stationary solution has finite moments of order $p$ if and only if $p < \alpha$,  
    that is,
    \begin{equation}
        \mathbb{E}\!\left[\|\x\|^p\right] < \infty \quad \text{iff } p < \alpha,
    \end{equation}
    and diverges for  $p \geq \alpha$.
    This result constitutes the $N$-dimensional extension of the classical scalar Kesten--Goldie theorem.
    
    \subsection{One-Dimensional Illustration}

    In the one-dimensional setting, the process reduces to
    \begin{equation}    \label{eq:apx_1d_kesten}
        x_{t+1} = \rho_t x_t + \sqrt{2\delta}\,\eta_t,
        \qquad \eta_t \overset{\textit{i.i.d.}}{\sim} \mathcal{N}(0,1),
    \end{equation}
    where $(\rho_t)$ is an i.i.d. sequence of random multipliers.  
    In this case, the Lyapunov exponent \eqref{eq:apx_lyapunov} simplifies to
    \begin{equation}
        \gamma = \mathbb{E}\!\left[\ln \rho_t\right].
    \end{equation}

    To obtain explicit numerical results, we must specify the distribution of $\rho_t$.  
    A convenient choice is to assume
    \(
        \ln \rho_t \sim \mathcal{N}(\ln\rho, \sigma_\rho^2),
    \)
    so that $\ln\rho_t$ is following a normal distribution with mean parameter $\ln\rho$.  
    Here, $\ln\rho$ being the expectation of $\ln \rho_t$ directly defines the Lyapunov exponent $ \gamma= \ln\rho$.

    Under this assumption, the power law exponent $\alpha$ is determined by the equation
    \begin{equation}
        \phi(\alpha) := \ln \mathbb{E}\!\left[\rho_t^\alpha\right] = 0.
    \end{equation}
    Evaluating this expression yields
    \begin{equation}
        \phi(\alpha) = \gamma \alpha + \tfrac{1}{2}\sigma_\rho^2 \alpha^2 = 0
        \quad \Longrightarrow \quad
        \alpha = -\,2\frac{\gamma}{\sigma_\rho^2}.
        \label{trhjyhn3whw}
    \end{equation}

    Therefore, as long as $\gamma < 0$ (ensuring average contraction), the system remains stable.  
    However, if the variance of the multiplicative noise satisfies $\sigma_\rho^2 > |\gamma|$,  
    the stationary distribution exists but its variance is infinite.  
    This illustrates how heavy-tailed behavior can arise even in a stable one-dimensional Kesten process.
    
    In the one-dimensional Kesten process \eqref{eq:apx_1d_kesten},  
    the emergence of power law tails can be understood in terms of spectral criticality.  
    In standard dynamical systems language, a system becomes critical when $\rho$ approaches the unit root, i.e. $\rho \to \pm 1$.  
    However, in a Kesten process, the stochastic fluctuations of $\rho_t$ imply that the process may occasionally cross the unit root,  
    since $1 > \mathbb{P}\!\left[|\rho_t| > 1\right] > 0$. The power law (\ref{dfnhtyq}) arises from the interplay between two opposing exponential processes
    \cite{Sornette1998,Sornette1998pre,Sornette2006}.
    Transient episodes where $|\rho_t|>1$ cause multiplicative (and thus exponential) amplification of $x_t$ over their duration $n$.
    However, such supercritical bursts occur with an exponentially decaying probability as $n$ increases.
    The balance between the exponential growth of $x_t$ during these rare intervals and their exponentially decreasing frequency produces an emergent power-law tail.
    In essence, the heavy tail reflects the statistics of transient supercritical phases that exponentially amplify $x_t$ but whose occurrence becomes exponentially rarer with increasing duration.

    \subsection{Normal Kesten Processes}

    In the $N$-dimensional case \eqref{eq:apx_kesten},  
    the same intuition applies: power law tails arise from fluctuations between sub-critical and super-critical regimes.  
    To make this precise, consider the decomposition
    \(
        \A_t = \PP \Lamb_t \PP^{-1},
    \)
    where $\PP$ is a unitary matrix i.e. $\PP^{-1}=\PP^\dag$;
    and $\Lamb_t = \mathrm{Diag}(\lambda_{i,t} \,|\, i=1,\dots,N)$,  
    with the eigenvalues $\lambda_{i,t}$ assumed i.i.d.  
    
    Because $\PP$ is unitary, the operator norm is invariant under conjugation:
\begin{equation}
    \Big\|\prod_{s=1}^t \A_s\Big\|
    = \Big\|\PP \Big(\prod_{s=1}^t \Lambda_s\Big) \PP^{-1}\Big\|
    = \Big\|\prod_{s=1}^t \Lambda_s\Big\|.
\end{equation}

Since the matrices $\Lambda_s$ are diagonal, they commute, and their product is also diagonal:
\begin{equation}
    \prod_{s=1}^t \Lambda_s 
    = \mathrm{Diag}\!\left(\prod_{s=1}^t \lambda_{1,s}, \dots, \prod_{s=1}^t \lambda_{N,s}\right).
\end{equation}
The operator norm of this product is the maximum absolute diagonal entry:
\begin{equation}
    \pi_t = \Big\|\prod_{s=1}^t \Lambda_s\Big\|
    = \max_{1 \le i \le N} \prod_{s=1}^t |\lambda_{i,s}|.
    \label{eq:pi_t}
\end{equation}
Defining
\[
    M_{i,t} := \sum_{s=1}^{t} \ln|\lambda_{i,s}| ,
\]
we have $\ln \pi_t = \max_i M_{i,t}$.
Hence $M_{i,t}$ represents the additive logarithmic growth
of the $i$-th eigenmode of the product, while
$\Pi_t$ (and its norm $\pi_t$) aggregates the maximum amplification
over all modes.  In the normal case, $M_{i,t}$ tracks mode-wise growth and
$M_{1,t}$ is a representative scalar process.

Taking the logarithm and dividing by $t$, we obtain
\begin{equation}
    \frac{1}{t}\ln \pi_t
    = \max_{1 \le i \le N}\, \frac{1}{t} \sum_{s=1}^t \ln |\lambda_{i,s}|  =  \frac{1}{t} \max_i M_{i,t}
    \label{eq:max_sum}~.
\end{equation}
Note that the maximum acts on the time-averaged logarithmic growth of each component $i$.  

Let us assume that $\{\lambda_{i,s}\}$ are i.i.d.\ across both $i$ and $s$, with finite mean logarithm,
$\mathbb{E}|\ln|\lambda_{1,1}|| < \infty$.  
For each fixed $i$, by the strong law of large numbers,
\begin{equation}
    \frac{1}{t}\sum_{s=1}^t \ln |\lambda_{i,s}|
    \xrightarrow[t \to \infty]{\text{a.s.}}
    \mathbb{E}\!\left[\ln|\lambda_{1,1}|\right].
    \label{eq:lln_component}
\end{equation}
Since this limit is identical for all $i$, the maximum in \eqref{eq:max_sum} converges almost surely to the same value:
for any $\varepsilon>0$, there exists $T$ such that for all $t>T$, each average lies within $\varepsilon$ of 
$\mathbb{E}[\ln|\lambda_{1,1}|]$, hence their maximum also does.  
Therefore,
\begin{equation}
    \lim_{t\to\infty} \frac{1}{t}\ln \pi_t
    = \mathbb{E}\!\left[\ln|\lambda_{1,1}|\right].
\end{equation}

The top Lyapunov exponent is thus
\begin{equation}
    \gamma
    = \lim_{t\to\infty} \frac{1}{t}\ln \Big\|\prod_{s=1}^t \A_s\Big\|
    = \mathbb{E}\!\left[\ln|\lambda_{1,1}|\right] =  \ln \rho ~.
\end{equation}
Criticality occurs when $\gamma = 0$, i.e.\ when $\mathbb{E}[\ln|\lambda_{1,1}|] = 0$.
Thus, the $N$-dimensional case reduces to the same structure as the scalar case,  
    and criticality occurs when $\gamma \to 0^{-}$.  
    When $\PP$ is constant and unitary, this corresponds directly to spectral criticality.
    \newline

    We now turn to the tail exponent. 
    From definition (\ref{eq:phi_defrtbt}), we have
    \begin{equation}
        \begin{split}
            &\phi(\alpha) = \lim_{t\to\infty} \frac{1}{t}
            \ln \mathbb{E}\!\left[e^{\alpha \ln \pi_t}\right], \\
            &\ln \pi_t = \max_{i=1,\dots,N} \sum_{s=1}^t \ln |\lambda_{i,s}|~,
        \end{split}
    \end{equation}
    where the expression for $\ln \pi_t$ derives from (\ref{eq:pi_t}). Since
    \begin{equation}
        M_{i,t} := \sum_{s=1}^t \ln |\lambda_{i,s}|
        \;\overset{t\to\infty}{\sim}\; \mathcal{N}(t \ln \rho,\, t\sigma_\rho^2),
    \end{equation}
    the problem reduces to analyzing the maximum of $N$ i.i.d. Gaussian random variables,  
    a classical extreme-value setting.  The following inequalities hold
    \begin{equation}
    \frac{1}{N}\sum_{i=1}^N e^{\alpha M_{i,t}}
    \;\le\;
    \max_{i} e^{\alpha M_{i,t}}
    \;=\;
    e^{\alpha \max_{i} M_{i,t}}
    \;\le\;
    \sum_{i=1}^N e^{\alpha M_{i,t}}.
    \label{ehtbgqf}
    \end{equation}
    From \eqref{ehtbgqf}, taking expectations and using that the $M_{i,t}$ are i.i.d., we obtain
    \begin{equation}
    \mathbb{E}\!\left[e^{\alpha M_{1,t}}\right]
    \;\le\;
    \mathbb{E}\!\left[e^{\alpha \max_{i} M_{i,t}}\right]
    \;\le\;
    N \,\mathbb{E}\!\left[e^{\alpha M_{1,t}}\right].
    \label{17qthgy2b}
    \end{equation}
    Hence,
    \begin{equation}
    \lim_{t\to\infty}\frac{1}{t}\ln \mathbb{E}[e^{\alpha M_{1,t}}]
        \;\leq\; \phi(\alpha) \;\leq\;
        \lim_{t\to\infty}\frac{1}{t}
        \Big[\ln \mathbb{E}[e^{\alpha M_{1,t}}] + \ln N\Big]~,
    \end{equation}
    leading to 
   \begin{equation}
       \phi(\alpha) = \lim_{t\to\infty}\frac{1}{t}
        \ln \mathbb{E}[e^{\alpha M_t}],
        ~~~ M_t \sim \mathcal{N}(t\ln \rho,\, t\sigma_\rho^2), 
   \end{equation}
   with $\ln \rho := \mathbb{E}\!\left[\ln|\lambda_{1,1}|\right]$ and
    $\sigma_\rho^2 := \mathrm{Var}\!\left[\ln|\lambda_{1,1}|\right]$.
    Therefore, provided $\sigma_\rho^2 < \infty$, the solution of $\phi(\alpha) = 0$ is
    \begin{equation}
        \alpha = -\,\frac{2\gamma}{\sigma_\rho^2}~,
        \label{wrynhy2bq}
    \end{equation}
    which has the same form as the one-dimensional expression (\ref{trhjyhn3whw}).
    
    Finally, consider the case where $\PP_t$ is also stochastic but remains unitary.  
    In this situation, $\A_t$ remains normal, and $\PP_t$ does not affect the norm of $\A_t$,  
    so that $\|\A_t\| = \|\Lamb_t\|$.  
    For a product  $\prod_{s=1}^t A_s$ of such matrices, we have
    \begin{equation}
    \pi_t = \left\|\prod_{s=1}^t A_s\right\|
    \;\le\;
    \prod_{s=1}^t \|A_s\|
    = \prod_{s=1}^t \|\Lambda_s\|,
    \label{2t4hyb20}
    \end{equation}
    where the inequality is the standard submultiplicativity of the operator norm.
    This provides an upper bound for the Lyapunov exponent:
    \begin{equation}
    \gamma \leq   \gamma_0 := \ln \rho ~.
    \end{equation}
    Thus, in the normal case, the scalar result (\ref{trhjyhn3whw}) and (\ref{wrynhy2bq}) provides an upper bound for the Lyapunov exponent $\gamma$.  

The submultiplicativity of the operator norm (inequality~(\ref{2t4hyb20}))
implies the upper bound for $\phi(\alpha)$  (\ref{eq:phi_defrtbt}),
\begin{equation}
    \phi(\alpha) \le  \phi_0(\alpha) 
     \label{wbgrwvq}
 \end{equation}
 where 
\begin{align}
\phi_0(\alpha)
&:= \lim_{t\to\infty}\frac{1}{t}
    \ln \mathbb{E}\!\left[e^{\alpha M_{1,t}}\right] \nonumber\\
&= \lim_{t\to\infty}
    \frac{1}{t}\,
    \ln
    \mathbb{E}\!\left[
        \exp\!\Big(
            \alpha
            \sum_{s=1}^{t}
            \ln\|A_s\|
        \Big)
    \right].
 \label{eq:phi_uppephi0r}
 \end{align}

    Since $\phi(\alpha)$ is convex, if $\alpha$ is the solution of $\phi(\alpha) = 0$,  
    then any $\alpha_0 > 0$ with $\phi_0(\alpha_0) = 0$ satisfies $\alpha_0 < \alpha$, where 
    $ \alpha_0 = -\,\frac{\gamma_0}{\sigma_\rho^2}$
    which provides a lower bound on the true tail exponent.
    The detailed proof of this result is as follows.
    The two functions $\phi,\phi_0:[0,\infty)\to\mathbb{R}$ are convex functions with
    \begin{equation}
        \phi(0)=\phi_0(0)=0,\; \phi'(0)=\gamma<0,\; \phi_0'(0)=\gamma_0\ge \gamma,
    \end{equation}
    and for all $\alpha\ge 0$ we have the pointwise bound (\ref{wbgrwvq}) with (\ref{eq:phi_uppephi0r}).
    By convexity and $\phi'(0)<0$, both $\phi$ and $\phi_0$ admit a \emph{unique} zero in $(0,\infty)$, which we denote by
    \begin{equation}
        \phi(\alpha)=0 \text{ at } \alpha>0,\qquad \phi_0(\alpha_0)=0 \text{ at } \alpha_0>0.
    \end{equation}
    Since $\phi_0(\alpha_0)=0$ and $\phi(\alpha)\le \phi_0(\alpha)$ for all $\alpha\ge 0$, we have
    \begin{equation}
        \phi(\alpha_0)\;\le\;\phi_0(\alpha_0)\;=\;0.
        \label{yjun4yjj}
    \end{equation}
    Because $\phi$ is convex with $\phi(0)=0$ and $\phi'(0)<0$, $\phi$ is strictly negative on a right-neighborhood of $0$ and then (by convexity) crosses $0$ exactly once at $\alpha>0$. 
    The inequality $\phi(\alpha_0)\le 0$ means that $\alpha_0$ lies at or to the left of the zero of $\phi$, hence $\alpha_0\le \alpha$.
    Moreover, if either $\gamma_0>\gamma$ or the bound is strict for some $\alpha>0$ (which is the case under the normal-vs-general comparison),
    then $\phi(\alpha_0)<0$ and the crossing of $\phi$ must occur strictly to the right of $\alpha_0$, yielding $\alpha_0<\alpha$.

    In summary, for normal random matrices $\A_t$,  
    the one-dimensional results yield an upper bound on the Lyapunov exponent  
    and a lower bound on the tail exponent. In other words, the one-dimensional Kesten process represents the
worst-case scenario of the multidimensional Kesten process with normal matrices in terms of stability and heavy-tailedness.

    \subsection{Non-Normal Kesten Processes}

    A matrix is called \emph{non-normal} if it cannot be diagonalized in a unitary basis,  
    that is, if $\PP_t^{-1} \neq \PP_t^\dag$.
    In this case, by the classical bound for diagonalizable matrices (see, e.g., \cite{HornJohnson2013}), with equality (and $\kappa = 1$) when the eigenbasis
   is orthogonal, the matrix norm satisfies the inequality
    \begin{equation}
        \|\A_t\| \leq  \rho_t\,\kappa_t ~,
        \qquad \kappa_t = \|\PP_t\|\,\|\PP_t^{-1}\|,
        \label{dhwgbwa}
    \end{equation}
    where $\kappa_t$ denotes the condition number of the eigenbasis transformation matrix $\PP_t$,
    and $\rho_t$ is the spectral radius of $\A_t$.
    Since the matrices $(\A_t)$ are i.i.d., the condition numbers $(\kappa_t)$ are also i.i.d.
    
    Taking products and using submultiplicativity,
\begin{equation}
\Big\|\prod_{s=1}^{t}A_s\Big\|
\;\le\;\prod_{s=1}^{t}\|A_s\|
\;\le\;\prod_{s=1}^{t}\big(\rho_s\,\kappa_s\big),
\label{yn3nhnbg}
\end{equation}
so that after logarithms and averaging,
\[
\frac{1}{t}\,\mathbb{E}\!\left[\ln\Big\|\prod_{s=1}^{t}A_s\Big\|\right]
\;\le\;\frac{1}{t}\,\mathbb{E}\!\left[\sum_{s=1}^{t}\ln\rho_s\right]
\;+\;\frac{1}{t}\,\mathbb{E}\!\left[\sum_{s=1}^{t}\ln\kappa_s\right].
\]
Letting $t\to\infty$ gives the Lyapunov exponent bound
    \begin{equation}
   \gamma \leq
        \gamma_0 = \lim_{t\to\infty}
        \left[ \frac{1}{t}\,\mathbb{E}\!\left[\sum_{s=1}^t \ln \rho_{s}\right]
            + \frac{1}{t}\,\mathbb{E}\!\left[\sum_{s=1}^t \ln \kappa_s\right]\right].
    \end{equation}
    By construction $\kappa_t \geq 1$, hence $\ln \kappa_t \geq 0$, and by the law of large numbers
    \begin{equation}    \label{eq:apx_lyapunov_non_normal}
       \gamma \leq  \gamma_0 = \ln \rho + \ln \kappa  
        \qquad {\rm with}~ \ln \kappa := \mathbb{E}[\ln \kappa_t] \geq 0.
    \end{equation}
    Thus, non-normality systematically increases the upper bound of the Lyapunov exponent compared with the normal case.
    \newline

    To quantify the effect on the tail exponent, we consider the log-condition number
    \begin{equation}
        \ln \kappa_t = \max_{i=1,\dots,N} \ln s_{i,t}
                    - \min_{j=1,\dots,N} \ln s_{j,t},
    \end{equation}
    where $(s_{i,t})$ are the singular values of $\PP_t$.  
    For simplicity, we assume that $(s_{i,t})$ are i.i.d. and independent of the eigenvalues $(\lambda_{i,t})$.  
This assumption can be justified on several theoretical and probabilistic grounds
within the framework of non-Hermitian random matrix theory, particularly for
ensembles such as the Ginibre ensemble \cite{ginibre1965statistical} (see below)
In a nutshell, the assumption of independence between singular values and eigenvalues
is a valid leading-order approximation in the ``far-from-criticality'' regime,
where eigenvalue magnitudes remain bounded away from unity and near-resonant
eigenvalue clusters are exponentially rare.
 The problem of estimating $\ln \kappa_t$ is then a classical problem of extreme-value theory (EVT).

Starting from the upper bound (\ref{yn3nhnbg}) on the matrix product norm, 
we introduce the cumulant generating function of the logarithmic growth:
\begin{equation}
    \phi_0(\alpha)
    := \lim_{t\to\infty} 
    \frac{1}{t}
    \ln
    \mathbb{E}\!\left[
        \exp\!\left(
            \alpha 
            \sum_{s=1}^{t}
            \ln\rho_s+  \sum_{s=1}^{t} \ln\kappa_s
        \right)
    \right].
    \label{eq:phi0_def}
\end{equation}
The lower bound $\alpha_0$ of the tail exponent $\alpha$ is given by (\ref{yjun4yjj}): $\phi_0(\alpha_0)=0$.
    Assuming independence between eigenvalues and condition numbers, we can treat the 
    two sums in (\ref{eq:phi0_def}) separately.  We have
    \begin{equation}
        \lim_{t\to\infty} \frac{1}{t}\,
        \ln \mathbb{E}\!\left[\exp\!\Big(\alpha \sum_{s=1}^t \ln\rho_{s}\Big)\right]
        = \alpha \ln \rho + \tfrac{1}{2}\sigma_\rho^2 \alpha^2.
    \end{equation}
    On the other hand,
    \begin{equation}
        \sum_{s=1}^t \ln \kappa_s \;\sim\; \mathcal{N}(t\ln \kappa,\, t\sigma_\kappa^2),
        \qquad t\to\infty,
    \end{equation}
    so that
    \begin{equation}
        \phi_0(\alpha) = \alpha (\ln \rho + \ln \kappa)
        + \tfrac{1}{2}\alpha^2 \big(\sigma_\rho^2 + \sigma_\kappa^2\big).
    \end{equation}
    The solution of $\phi_0(\alpha_0) = 0$ is therefore
    \begin{equation} \label{eq:apx_tail_non_normal}    
        \alpha_0 = -\,\frac{\gamma_0}{\sigma_\rho^2 + \sigma_\kappa^2} \le \alpha.
    \end{equation}
    This result holds for any distribution of the singular values $(s_{i,t})$,  
    provided that $\mathrm{Var}(\ln s_{i,t}) < \infty$.
    \newline

    In summary, non-normality of the matrices $\A_t$ increases the upper bound of the Lyapunov exponent  
    and decreases the lower bound of the tail exponent.  
    Consequently, non-normality enlarges the parameter region where the system approaches instability ($\gamma \geq 0$),  
    and can simultaneously  enhances the heaviness of the tail of the stationary distribution.

    \subsection{Eigenvalue--Eigenvector Independence Assumption}

  In this section, we present generic results based on the Ginibre ensemble, which will later support qualitative arguments justifying the assumption that the eigenvectors and eigenvalues of the system can be treated as independent when the spectral radius remains well below unity (i.e., far from the critical boundary).
    \newline

    Let \(\A\in\mathbb R^{N\times N}\) denote a (real) Ginibre matrix \cite{ginibre1965statistical} with i.i.d. entries
    \(
        (\A)_{ij}\sim \mathcal N \big(0,\;1/\sqrt{N}\big)
    \),
    so that the empirical spectral radius is \(O(1)\) and the eigenvalue cloud obeys the circular law (with its support being approximately the unit disk).  We denote the eigenvalue-eigenvector decomposition (when diagonalizable)
    \(
    \A = \PP \Lamb \PP^{-1},\, \Lamb = \text{Diag}(\lambda_1,\dots,\lambda_N),
    \)
    with (right) eigenvectors \(\rr_i\) (columns of \(\PP\)) and left eigenvectors \(\el_i^\dag\) (rows of \(\PP^{-1}\)),
    normalized so that \(\el_i ^\dag \rr_i = 1\).
    The local (or eigenpair) condition / overlap is defined as
    \begin{equation}
        \mathcal O_i \;=\; \|\rr_i\|\,\|\el_i\|,
    \end{equation}
    and the global eigenbasis condition number is
    \begin{equation}
        \kappa(P_N) \;=\; \|P_N\|\,\|P_N^{-1}\| \;\gtrsim\; \max_{1\le i\le N} \mathcal O_i.
    \end{equation}
Below, we summarize the key statistical properties of the ensemble that are essential for the subsequent analysis.
In particular, the spectral and overlap properties are standard in the random matrix literature on non-Hermitian matrices \cite{chalker1998eigenvector,bourgade2019distribution}.
 
    \begin{itemize}[label={}, leftmargin=0pt]
    \item \emph{Circular law and bulk regime.}
    As \(N\to\infty\), the eigenvalues fill the unit disk (circular law).
    Throughout this work, we consider eigenvalues conditioned to lie well inside the spectral boundary,
    i.e.\ \(|\lambda_i|<1-\varepsilon\) for some fixed \(\varepsilon>0\). 
    This is the ``far from spectral criticality'' regime:
    spectral radii are bounded away from unity and the leading eigenvalue magnitudes do not approach instability thresholds.

    \item \emph{Typical eigenvector non-orthogonality (overlaps).} 
    For the complex Ginibre ensemble, the diagonal overlap (for an eigenvalue conditioned at \(z\) in the bulk) has expectation of order
    \begin{equation}
        \mathbb E\big[\mathcal O_i \mid \lambda_i=z\big] \asymp \frac{N}{1-|z|^2}\quad(|z|<1).
    \end{equation}
    Thus overlaps scale linearly with \(N\) (and blow up as \(|z|\to 1\)).
    This is the Chalker--Mehlig scaling later made rigorous by subsequent authors;
    it is the principal reason non-normal effects grow with system size in the bulk.
    Intuitively, the factor \(1/(1-|z|^2)\) encodes the reduced local spectral stability near the boundary while the factor \(N\) is a combinatorial / density effect of the many degrees of freedom \cite{chalker1998eigenvector,bourgade2019distribution}.
    
    \item \emph{Heavy tails and rare large overlaps.}
    Overlaps have heavy-tailed fluctuations: the distribution of \(\mathcal O_i\) (properly rescaled) admits heavy tails / inverse-Gamma type limits in the bulk.
    Consequently the maximum overlap (hence \(\kappa(\PP)\)) is dominated by rare large events (near-colliding eigenvalues, tight local clusters),
    not just by the typical value above.  This heavy-tailed character is important when computing extreme rare-event contributions to Lyapunov/tail exponents.

    \item \emph{Local spacing dependence.} 
    When an eigenvalue \(\lambda_i\) has a nearby eigenvalue at distance \(\Delta\),
    the corresponding overlaps can scale like \((N\Delta^2)^{-1}\):
    small nearest-neighbor gaps produce large overlaps. 
    In practice this means that configurations with small eigenvalue gaps (even if rare) strongly increase \(\kappa(\PP)\).
    Thus the geometry of the eigenvalue cloud (positions and spacings) directly controls eigenvector conditioning.
    \end{itemize}

  In several derivations of Kesten dynamics throughout this paper, we assume that 
the statistical properties of the eigenbasis (singular values and eigenvectors) 
are approximately independent from those of the eigenvalue set. 
This approximation is well controlled when the system is \emph{far from criticality}, 
that is, when all eigenvalues satisfy $|\lambda_i| < 1 - \varepsilon$ uniformly 
and no systematically small eigenvalue gaps occur. 
This assumption can be justified on several theoretical and probabilistic grounds, 
supported by heuristic reasoning.

    \begin{itemize}[label={}, leftmargin=0pt]
      \item \emph{Bi-unitarily invariance:}
    For the complex Ginibre ensemble, the distribution of matrices is exactly 
bi--unitarily invariant for any $N$, meaning that
$A \stackrel{d}{=} U A V$ for all $U,V\in \mathrm{U}(N)$.
This invariance ensures that the singular vectors are uniformly distributed 
and statistically independent of the singular values.
However, this bi-invariance does \emph{not} imply that the eigenvectors and 
eigenvalues are independent: at finite $N$, their joint density involves 
correlations through the non--orthogonality matrix~$G$  \cite{chalker1998eigenvector}.
In the large--$N$ limit, these correlations vanish in the bulk of the spectrum, 
so that eigenvectors and eigenvalues become asymptotically independent.
For non--Gaussian i.i.d.\ ensembles, the same asymptotic factorization holds 
by universality, but independence is no longer exact at finite~$N$.
        \item \emph{Bulk decorrelation:}
        for Ginibre matrices, the eigenvectors (left/right) are,
        at leading order and in distribution, approximately isotropic objects whose directional statistics are weakly dependent on the macroscopic eigenvalue location provided the location is in the bulk
        (i.e.\ at distance \(>\!\varepsilon\) from the spectral edge). 
        Thus, conditioning on \(\lambda_i=z\) with \(|z|<1-\varepsilon\) modifies eigenvector statistics only mildly at \(O(1)\) level while the typical overlaps scale like \(N\),
        which is a separation of scales that allows approximate factorization in many averaged calculations.

        \item \emph{Absence of near-resonances:}
        the main failure mode of independence is clustering / near-collisions of eigenvalues:
        if two eigenvalues are atypically close, eigenvectors associated with them are strongly coupled and independence breaks down.
        By conditioning away from small spacings i.e. assuming a minimal spacing \(\Delta_{\min}\) not smaller than \(O(N^{-1/2+\eta})\) for some \(\eta>0\);
        these rare pathological configurations are excluded and the independence approximation becomes valid for the dominant contributions that we study.

        \item \emph{Controlled error in Kesten averages:}
        in the Kesten-product formulas the eigenvalue contribution enters multiplicatively via \(\prod_t |\lambda_{i,t}|\) while non-normal amplification enters via multiplicative factors related to \(\kappa(P_t)\).
        When eigenvalues stay uniformly subcritical (\(|\lambda|<1-\varepsilon\)) the product of spectral moduli decays and the leading corrections coming from eigenvector-eigenvalue correlations manifest as finite \(O(1)\) modifications to Lyapunov or tail exponents;
        they do not change the scaling exponents unless the system approaches spectral criticality (where small changes can flip the sign of the Lyapunov exponent).
        This is precisely why assuming independence is safe in the far from criticality regime used in our main results. 
    \end{itemize}
The independence assumption provides accurate 
leading-order predictions for the Lyapunov and tail exponents, as well as for the 
scaling of non-normal amplification, while rare violations can be treated separately 
as extreme events.

    \subsection{Synthesis}

    We have introduced Kesten processes, a class of stochastic recursions 
    driven by random multiplicative dynamics and additive noise.  
    In the one-dimensional setting, such processes exhibit stationary distributions with heavy-tailed 
    power law behavior, provided the Lyapunov exponent $\gamma$ is negative (ensuring average contraction) 
    but the local dynamics occasionally exceed the unit root.  
    The tail exponent $\alpha$ is then determined by the balance between contraction and fluctuations, 
    capturing how moments of the stationary solution may diverge.

    Extending this picture to higher dimensions reveals the key mechanism is unchanged: 
    criticality emerges when random fluctuations intermittently push the system across the spectral 
    stability boundary.  
    For normal random matrices, the Lyapunov exponent and the power law exponent can be bounded 
    using the one-dimensional theory, linking criticality directly to the eigenvalue spectrum.  
    This corresponds to the familiar notion of \emph{spectral criticality} in natural systems, 
    where behavior changes qualitatively as eigenvalues approach the unit circle.

    However, when the random matrices are non-normal, a new phenomenon arises.  
    Because the condition number of the eigenbasis amplifies fluctuations, non-normality 
    increases the effective Lyapunov exponent and decreases the tail exponent, 
    thereby enlarging the region of parameter space where the system behaves as if it were critical.  
    In other words, even when the spectrum indicates stability, non-normality can drive the process 
    closer to true criticality and produce heavier-tailed stationary distributions.  
    This highlights non-normality as a key mechanism that can mimic or enhance critical behavior 
    in stochastic dynamical systems.

    We presented some generic results associated to the Ginibre ensemble,
    allowing us to use later in this paper, that conditionally for the matrix spectral radius to be ``fare'' from the unity,
    the eigenbasis and eigenvectors of the random matrices are independent.

    In the next section, we turn to concrete applications.  
    By working through explicit examples, we will illustrate how the theoretical bounds derived here 
    manifest in practice, and how non-normality alters the effective stability and tail properties 
    of Kesten processes in applied settings.


    \section{Two Dimensional Illustration}


    The general theory developed in the previous section applies to arbitrary dimension, 
    but it can be difficult to develop intuition.  
    To gain more insight, we now focus on the two-dimensional case.  
    This setting is simple enough to allow explicit calculations, 
    while still capturing the essential role of non-normality in modifying stability and tail behavior.  

    We proceed step by step: first by analyzing a tractable special form of random matrices 
    (where two-step products become diagonal), then by studying how the Lyapunov exponent and 
    the tail exponent can be computed exactly, and finally by comparing these results 
    with the general bounds derived earlier.  
    This two-dimensional illustration serves both as a consistency check 
    and as a bridge toward more complex applications presented later.

    \subsection{A First Example}

    To gain intuition about the effect of non-normality in higher dimensions, let us start with a 
    two-dimensional example. We consider matrices of the form
    \begin{equation}    \label{eq:apx_simp_2d}
        \A_t = \rho_t
        \begin{pmatrix}
            0 & z_t \\
            z_t^{-1} & 0
        \end{pmatrix},
        \qquad z_t = \frac{s_{1,t}}{s_{2,t}},
    \end{equation}
    where $\rho_t$ sets the spectral radius, and $s_{1,t}, s_{2,t}$ are the singular values of the 
    eigenbasis transformation matrix $\PP_t$. We assume that the $(s_{i,t})$ are 
    independent and identically distributed (i.i.d.) and also independent of the i.i.d. multipliers $(\rho_t)$.
    \newline

    Our goal is to compute explicitly the Lyapunov exponent $\gamma$ and the tail exponent $\alpha$. 
    We begin with the product norm $\pi_t$ defined in~\eqref{eq:apx_matrix_prod}. 
    A key simplification arises if we look at two-step products:
    \begin{equation}
        \begin{split}
            \A^{(2)}_t &= \A_{2t}\A_{2t-1}  \\
            &= \rho_{2t}\rho_{2t-1}
            \begin{pmatrix}
                z_t^{(2)} & 0 \\
                0 & (z_t^{(2)})^{-1}
            \end{pmatrix},
            \quad z_t^{(2)} = \frac{z_{2t}}{z_{2t-1}}.
        \end{split}
    \end{equation}
    Thus, every two-step product is diagonal, which makes the analysis tractable as it reduces to the one-dimension set-up. The product norm over 
    $2t$ steps reads
    \begin{equation}
        \begin{split}
            \pi_{2t} 
            &= \left\|\prod_{s=1}^t \A^{(2)}_s \right\| \\
            &= \left(\prod_{s=1}^{2t} |\rho_s|\right)
            \max\!\left(Z_t, Z_t^{-1}\right),
            \quad Z_t = \prod_{s=1}^t z_s^{(2)}.
        \end{split}
    \end{equation}
    Taking logarithms and using monotonicity of $\ln(\cdot)$ gives
    \begin{equation}
        \ln \pi_{2t} 
        = \sum_{s=1}^{2t} \ln |\rho_s| + \big|\ln Z_t\big|.
    \end{equation}

    By the law of large numbers, the Lyapunov exponent is
    \begin{equation}
        \gamma = \ln\rho + \tfrac{1}{2}\big|\mathbb{E}[\ln z^{(2)}_1]\big|,
        \qquad \ln\rho := \mathbb{E}[\ln|\rho_1|].
    \end{equation}
    But since $z^{(2)}_t = z_{2t}/z_{2t-1}$ and the $z_i$ are i.i.d., we have
    $\mathbb{E}[\ln z^{(2)}_1] = \mathbb{E}[\ln z_2 - \ln z_1] = 0$. Hence the Lyapunov exponent simplifies to
    \begin{equation} \label{eq:apx_lyapunov_2d}
        \gamma = \ln\rho.
    \end{equation}
    In other words, in this specific construction, the non-normality cancels out on average, and the Lyapunov exponent 
    is determined solely by the spectral radius.
    \newline

    The tail behavior, however, is more subtle. It depends on the distribution of $\ln z_t^{(2)}$. 
    If $\mathrm{Var}[\ln z_t^{(2)}] < \infty$, then by the Central Limit Theorem,
    \begin{equation}
        \ln Z_t \sim \mathcal{N}(0, 2t\sigma_z^2),
        \qquad \sigma_z^2 := \mathrm{Var}[\ln z_1].
    \end{equation}
    At the same time,
    \begin{equation}
        \sum_{s=1}^{2t} \ln |\rho_s| 
        \sim \mathcal{N}(2t\ln\rho,\, 2t\sigma_\rho^2),
        \qquad \sigma_\rho^2 := \mathrm{Var}[\ln|\rho_1|].
    \end{equation}
    Therefore, the logarithm of the product norm is approximately Gaussian:
    \begin{equation}
        \ln\pi_{2t} \sim 
        \mathcal{N}\big(2t\ln\rho,\; 2t(\sigma_\rho^2 + \sigma_z^2)\big).
    \end{equation}
    It follows that the tail exponent is given by the solution of $ \phi(\alpha)=0$ with $\phi(\alpha)$ given by (\ref{eq:phi_defrtbt}), which yields
    \begin{equation} \label{eq:apx_tail_2d}
        \alpha = -\frac{2\gamma}{\sigma_\rho^2 + \sigma_z^2}.
    \end{equation}
    Thus, in this case, non-normality does not affect stability (since $\gamma$ remains unchanged), 
    but it does decrease the tail exponent and therefore increases the heaviness of the stationary distribution.
   This result shows that the geometry of the eigenvectors by itself can produce and enhance fat-tailed stationary distributions.
Remarkably, even when transient supercritical excursions are absent ($\sigma_\rho =0$), 
 the non-normal amplification mechanism alone gives rise to power-law tails characterized by $   \alpha = -\frac{2\gamma}{ \sigma_z^2}$.
 
    Finally, we observe that the log-condition number is
    \begin{equation}
        \ln\kappa_t = |\ln z_t|,
    \end{equation}
    which implies $\sigma_z^2 \geq \sigma_\kappa^2$. Hence, the explicit solutions
    for the Lyapunov exponent~\eqref{eq:apx_lyapunov_2d} and the tail exponent~\eqref{eq:apx_tail_2d}
    are consistent with the general bounds derived in \eqref{eq:apx_lyapunov_non_normal} 
    and \eqref{eq:apx_tail_non_normal}.

    \subsection{General Case}

    In the previous subsection, we have considered a special $2\times 2$ construction 
    for which the Lyapunov exponent turned out to be independent of non-normality.  
    To understand the situation more generally, recall from~\cite{troude2025} 
    that any non-normal $2\times 2$ matrix can be written in the form
    \begin{equation}    \label{eq:apx_2d_matrix}
        \A = \tfrac{1}{2}\,\U
        \begin{pmatrix}
            \lambda_1 + \lambda_2 & z(\lambda_1 - \lambda_2) \\
            z^{-1}(\lambda_1 - \lambda_2) & \lambda_1 + \lambda_2
        \end{pmatrix} \U^\dag,
    \end{equation}
    where $\lambda_1,\lambda_2$ are the eigenvalues of $\A$,
    $z = s_1/s_2$ is the ratio of the singular values of the eigenbasis transformation matrix $\PP$, 
    and $\U$ is a unitary matrix.

    For real $2\times 2$ matrices, the eigenvalues are either both real or appear as complex conjugates.  
    In both cases, the spectral radius is given by
    \begin{equation}
        \begin{split}
            &\rho = \max(|\lambda_1|,|\lambda_2|) \quad \text{(real case)}, \\
            &\rho = |\lambda_1| = |\lambda_2| \quad \text{(complex conjugate case)}.
        \end{split}
    \end{equation}

    \begin{itemize}[label={}, leftmargin=0pt]
        \item \emph{Case 1: real eigenvalues.}
        When the eigenvalues are real, it is convenient to introduce the relative spectral distance
        \begin{equation}
            \delta = \frac{|\lambda_1|-|\lambda_2|}{|\lambda_1|+|\lambda_2|} \in [0,1],
        \end{equation}
        which quantifies how far apart the eigenvalues are compared to their average size.  
        With this notation, the matrix can be expressed as
        \begin{equation}
            \A = \rho \,\U
            \begin{pmatrix}
                \delta & (1-\delta)z \\
                (1-\delta)z^{-1} & \delta
            \end{pmatrix} \U^\dag.
            \label{twrhy2bq}
        \end{equation}
        
        \item\emph{Case 2: complex conjugate eigenvalues.}
        If the eigenvalues are complex conjugates, one can instead define
        \begin{equation}
            \delta = \frac{\Re(\lambda_1)}{|\lambda_1|} \in [-1,1],
        \end{equation}
        which captures the ratio between the real and modulus parts of the eigenvalue.  
        The matrix then takes the form
        \begin{equation}
            \A = \rho \,\U
            \begin{pmatrix}
                \delta & i\sqrt{1-\delta^2}\,z \\
                i\sqrt{1-\delta^2}\,z^{-1} & \delta
            \end{pmatrix} \U^\dag~.
            \label{twrhy2bqee}
        \end{equation}
    \end{itemize}

    In both situations, the parameters $\rho$, $\delta$,
    and $z$ jointly control the stability and tail properties of the corresponding Kesten process.
    The parameter $\rho$ governs the spectral radius (and thus the baseline stability condition), 
    $\delta$ measures the distance to spectral degeneracy, 
    and $z$ encodes the non-normality of the eigenbasis.  

    It is particularly informative to study the limiting regimes $\delta \approx 0$ 
    (close to spectral degeneracy, where non-normality plays the strongest role), 
    and $1-\delta \approx 0$ (where the system is nearly normal).  
   In the next subsections, we analyze how these different regimes affect 
    the Lyapunov exponent and the tail exponent.
    
    \subsection{Approximation for Large Non-Normality}

    For the general two-dimensional case (\ref{twrhy2bq}) and (\ref{twrhy2bqee}), let us study
    how non-normality modifies the Lyapunov and tail exponents.  
    To make analytical progress, we consider the regime where the parameter $\delta$ 
    is small (i.e.~close to spectral degeneracy).  
    In this setting, perturbation theory provides a useful approximation by expanding 
    the dynamics to first order in $\delta$.
    \newline

    \begin{itemize}[label={}, leftmargin=0pt]
        \item \emph{Case 1: real eigenvalues.}
         When the eigenvalues are real, the matrix can be expressed as
        \begin{equation}
            \A_t = \rho_t \Big( A_{0,t} + \delta(\I - A_{0,t}) \Big),
            \qquad 
            A_{0,t} =
            \begin{pmatrix}
                0 & z_t \\
                z_t^{-1} & 0
            \end{pmatrix}.
        \end{equation}
        Expanding the product over $2t$ steps gives
        \begin{widetext}
            \begin{equation}
                \PPi_{2t} = \prod_{s=1}^{2t} \A_s 
                = \Big(\prod_{s=1}^{2t} \rho_s\Big)
                \prod_{s=1}^{2t} \Big[ A_{0,s} + \delta(\I - A_{0,s}) \Big] 
                = \Big(\prod_{s=1}^{2t} \rho_s\Big)
                \left[
                        \prod_{s=1}^{2t} A_{0,s}
                        + \delta \sum_{r=1}^{2t}
                        \Big(\prod_{s=1}^{r-1} A_{0,s}\Big)
                        (\I - A_{0,r})
                        \Big(\prod_{s=r+1}^{2t} A_{0,s}\Big)
                \right] + \mathcal{O}(\delta^2).
            \end{equation}
        \end{widetext}
        After simplification, this reduces to
        \begin{equation}
            \PPi_{2t} = \Big(\prod_{s=1}^{2t} \rho_s\Big)
                        \Big(\prod_{s=1}^{2t} A_{0,s}\Big)
                        \Big[ \I - 2t\delta(\I - \B_t) \Big] + \mathcal{O}(\delta^2),
        \end{equation}
        where
        \begin{equation}
            \begin{split}
                &\qquad\qquad \B_t = 
                \begin{pmatrix}
                    0 & k^+_{2t} \\
                    k^-_t & 0
                \end{pmatrix}, \\
                &k^+_t = \frac{1}{t}\sum_{s=1}^t z_s^{(-1)^s}
                ,\quad 
                k^-_t = \frac{1}{2t}\sum_{s=1}^{2t} z_s^{(-1)^{s+1}}.
            \end{split}
        \end{equation}
        We note that the product of the $A_{0,s}$ terms simplifies to a diagonal matrix.  
        In particular, defining
        \begin{equation}
        Z_t := \prod_{s=1}^t z_s,
        \end{equation}
        we have
        \begin{equation}
        \prod_{s=1}^{2t} A_{0,s} = \mathrm{Diag}\!\left(Z_t,\,Z_t^{-1}\right).
        \tag{53}
        \end{equation}
        We thus obtain
        \begin{equation}
            \ln \pi_{2t} = \sum_{s=1}^{2t} \ln|\rho_s| + |\ln Z_t| - 2t\delta + \mathcal{O}(\delta^2).
        \end{equation}
        Therefore, in the large-$t$ limit,
        \begin{equation}
            \ln \pi_{2t} \sim \mathcal{N}\!\Big(2t(\ln\rho - \delta),\; 2t(\sigma_\rho^2 + \sigma_z^2)\Big)
                            + \mathcal{O}(\delta^2).
        \end{equation}
        This implies the approximations
        \begin{equation}
            \gamma = \ln\rho - \delta + \mathcal{O}(\delta^2),
            \qquad
            \alpha = -\frac{2\gamma}{\sigma_\rho^2 + \sigma_z^2} + \mathcal{O}(\delta^2).
        \end{equation}
        Thus, to first order in $\delta$, the Lyapunov exponent decreases linearly with the degree of spectral degeneracy, 
        while the tail exponent retains the same form as in the $\delta=0$ case. The correction to the tail exponent $\alpha$
       appears only at second order in $\delta$.

        \item \emph{Case 2: complex conjugate eigenvalues.}
        When the eigenvalues are complex conjugates, the expansion reads
        \begin{equation}
            \A_t = i\rho_t \Big( A_{0,t} - i\delta\I \Big) + \mathcal{O}(\delta^2).
        \end{equation}
        The product then becomes
        \begin{equation}
            \begin{split}
                \PPi_{2t} &= \prod_{s=1}^{2t} \A_s \\
                &= i^{2t}\Big(\prod_{s=1}^{2t} \rho_s\Big)
                \prod_{s=1}^{2t} \Big(A_{0,s} - i\delta\I\Big) + \mathcal{O}(\delta^2) \\
                &= i^{2t}\Big(\prod_{s=1}^{2t} \rho_s\Big)
                \Big(\prod_{s=1}^{2t} A_{0,s}\Big)\Big[ \I - 2i\delta t \B_t \Big]
                + \mathcal{O}(\delta^2).
            \end{split}
        \end{equation}
        Consequently, the logarithm of the product norm is
        \begin{equation}
            \ln\pi_{2t} = \sum_{s=1}^{2t} \ln|\rho_s| + |\ln Z_t| + \mathcal{O}(\delta^2).
        \end{equation}
        In this case, the first-order $\delta$-correction cancels, so the Lyapunov and tail exponents 
        are unchanged at order $\delta$:
        \begin{equation}
            \gamma = \ln\rho + \mathcal{O}(\delta^2), 
            \qquad 
            \alpha = -\frac{2\gamma}{\sigma_\rho^2 + \sigma_z^2} + \mathcal{O}(\delta^2).
        \end{equation}
    \end{itemize}

    To summarize, when $\delta$ is small, perturbation theory shows that:
    \begin{itemize}
        \item For real eigenvalues, non-normality decreases the Lyapunov exponent linearly in the degree $\delta$ of spectrum degeneracy, 
            slightly pushing the system closer to instability.
        \item For complex conjugate eigenvalues, the first-order correction vanishes, 
            and non-normality affects the exponents only at order $\delta^2$.
    \end{itemize}
    These results connect smoothly with the exactly solvable case $\delta=0$, 
    and illustrate how the influence of non-normality depends sensitively 
    on the spectral structure of the underlying matrices.

    \subsection{Non-Normal Reinjection via Rotation}

    So far, we have mostly considered the case where the rotation matrix $\U$ in 
    \eqref{eq:apx_2d_matrix} is fixed.  
    In that setting, non-normality affects the distribution of growth rates but 
    its influence on the Lyapunov and tail exponents iss limited.  
    In the general case, however, $\U_t$ can itself be stochastic and time dependent.  
    This introduces an additional mechanism: the rotation can continually 
    \emph{reinject the dynamics} into the direction of the most expanding eigenvector.  
    As a consequence, non-normality can substantially amplify the growth of trajectories 
    and modify both the Lyapunov exponent and the tail exponent.
    \newline

    To illustrate this phenomenon, let us focus on the simplest case of real eigenvalues
    with $\delta=0$.  
    The matrices then take the form
    \begin{equation}
        \begin{split}
            &\A_t = \rho_t \U(\theta_t)
            \begin{pmatrix}
                0 & z_t \\
                z_t^{-1} & 0
            \end{pmatrix}
            \U(\theta_t)^\dag, \\
            &\U(\theta) =
            \begin{pmatrix}
                \cos\theta & -\sin\theta \\
                \sin\theta & \cos\theta
            \end{pmatrix}.
        \end{split}
        \label{yjnhyn2}
    \end{equation}
    A useful identity is
    \begin{equation}
        \U(\theta)^\dag \U(\phi)
        = \cos(\theta+\phi)\I + \sin(\theta-\phi)
        \begin{pmatrix}
            0 & 1 \\
            -1 & 0
        \end{pmatrix},
        \label{trjun3hwg}
    \end{equation}
    which will allow us to keep track of how rotations couple across time.
   
  The product of two consecutive matrices 
    \begin{equation}
        \A^{(2)}_t \;=\; \A_{2t} \A_{2t-1}
        \label{63qgga}
    \end{equation}
    can be obtained explicitly using the identity (\ref{trjun3hwg}) to obtain
    \begin{equation}
        \begin{split}
            \A^{(2)}_t = \rho_{2t}\rho_{2t-1}\,\U(\theta_{2t})
            \Bigl[
                &\cos(\theta_{2t}+\theta_{2t-1})\,\D^{(2)}_t \\
                &+ \sin(\theta_{2t}-\theta_{2t-1})\,\K^{(2)}_t
            \Bigr]
            \U(\theta_{2t-1})^\dag,
        \end{split}
        \label{trwn3ws}
    \end{equation}
    \begin{equation}
        \text{where}\;
        \D^{(2)}_t =
        \begin{pmatrix}
            z_t^{(2)} & 0 \\
            0 & (z_t^{(2)})^{-1}
        \end{pmatrix},
        \qquad
        z_t^{(2)} = \frac{z_{2t}}{z_{2t-1}},
    \end{equation}
    \begin{equation}
        \text{and}\;
        \K^{(2)}_t =
        \begin{pmatrix}
            0 & -k^{(2)} \\
            1/k^{(2)} & 0
        \end{pmatrix},
        \qquad
        k^{(2)} = z_{2t}z_{2t-1}.
    \end{equation}
    Unlike the purely diagonal case studied before, the product $\A^{(2)}_t$ 
    is now a combination of a diagonal contribution $\D^{(2)}_t$ (which tends to average out non-normality) 
    and an anti-diagonal contribution $\K^{(2)}_t$ (which can reinforce it).  
    This constructive ``reinjection'' is the key new mechanism.
    \newline
    
   The impact of the anti-diagonal contribution $\K^{(2)}_t$ can be quantified by noting the following.
  Let us consider a matrix
    \begin{equation}
    \begin{split}
        &\A_t \;=\; \rho_t\, \U(\theta_t)\,\B(z_t)\,\U(\theta_t)^\dagger, \\
        &\B(z):=\begin{pmatrix}0&z\\ z^{-1}&0\end{pmatrix},
        \qquad
        \U(\theta)=\begin{pmatrix}\cos\theta&-\sin\theta\\ \sin\theta&\cos\theta\end{pmatrix}.
    \end{split}
    \label{wtrh2ty}
    \end{equation}
    Consider a two-step block \(\A^{(2)}_s:=\A_{2s}\A_{2s-1}\). Using unitarity,
    \begin{equation}
    \small
    \A^{(2)}_s
    = \rho_{2s}\rho_{2s-1}\,
    \U(\theta_{2s})\,
    \Bigl[\, \B(z_{2s})\,\R_{2s,2s-1}\,\B(z_{2s-1}) \Bigr]\,
    \U(\theta_{2s-1})^\dagger,
    \end{equation}
    where
    \begin{equation}
        \begin{split}
            \R_{2s,2s-1} \;&=\; \U(\theta_{2s})^\dagger \U(\theta_{2s-1}) \\
            &= \cos(\theta_{2s}+\theta_{2s-1})\,\I \;+\; \sin(\theta_{2s}-\theta_{2s-1})
            \begin{pmatrix} 0 & 1 \\ -1 & 0 \end{pmatrix},
        \end{split}
        \label{6th2t4h2}
    \end{equation}
    This allows us to obtain the two identities (checked by direct multiplication):
    \begin{equation}
        \begin{split}
            &\D^{(2)}(z,w) := \B(z)\,\I\,\B(w) \;=\; 
            \begin{pmatrix}
                z/w & 0 \\[2pt]
                0 & (z/w)^{-1}
            \end{pmatrix} ,
            \\
            &\K^{(2)}(z,w) := 
            \B(z)\,\Q\,\B(w) \;=\;
            \begin{pmatrix}
                0 & -zw \\[2pt]
                (zw)^{-1} & 0
            \end{pmatrix}.
        \end{split}
    \end{equation}
    Hence
    \begin{equation}
        \begin{split}
            \B(z_{2s})\,\R_{2s,2s-1}\,\B(z_{2s-1})
            &=
            \cos(\theta_{2s}+\theta_{2s-1})\,\D^{(2)}_s \\
            &\; +
            \sin(\theta_{2s}-\theta_{2s-1})\,\K^{(2)}_s ~.
        \end{split}
    \end{equation}
   
    Operator norms are invariant under unitary rotations before or after a matrix, so
    \begin{equation}
        \begin{split}
            \bigl\|\A^{(2)}_s\bigr\|
            = \bigl|\rho_{2s}\rho_{2s-1}\bigr|\;
            &\Bigl\|\cos(\theta_{2s}+\theta_{2s-1})\,\D^{(2)}_s \\
            &+\sin(\theta_{2s}-\theta_{2s-1})\,\K^{(2)}_s\Bigr\|.
        \end{split}
    \end{equation}
    In the non-normal (anti-diagonal)-dominated regime, we keep the \(\K^{(2)}\)-term,
    \begin{equation}
        \begin{split}
            \bigl\|\A^{(2)}_s\bigr\|
            &\approx
            \bigl|\rho_{2s}\rho_{2s-1}\bigr|\;\bigl|\sin(\theta_{2s}-\theta_{2s-1})\bigr|\;\bigl\|\K^{(2)}_s\bigr\| \\
            &=
            \bigl|\rho_{2s}\rho_{2s-1}\bigr|\;\bigl|\sin(\theta_{2s}-\theta_{2s-1})\bigr|\;\max\!\bigl(k^{(2)}_s,(k^{(2)}_s)^{-1}\bigr).
        \end{split}
    \end{equation}
    Multiplying the \(t\) two-step blocks and using submultiplicativity,
    \begin{equation}
        \begin{split}
            \pi_{2t}
            &=\Bigl\|\prod_{s=1}^{2t}A_s\Bigr\|
            =\Bigl\|\prod_{s=1}^{t}A^{(2)}_s\Bigr\| \\
            &\approx\;
            \Bigl(\prod_{s=1}^{2t}|\rho_s|\Bigr)\,
            \Bigl(\prod_{s=1}^{t}|\sin(\theta_{2s}-\theta_{2s-1})|\Bigr)\,
            \max\!\bigl(K_t,K_t^{-1}\bigr), \\
            \text{with}\; 
            &
            K_t:=\prod_{s=1}^{t}k^{(2)}_s=\prod_{s=1}^{2t}z_s.
        \end{split}
    \label{6qtgtgyb4}
    \end{equation}
    Taking logarithms and writing \(\ln\kappa_s:=|\ln z_s|\) so that
    \(\sum_{s=1}^{t}\ln\max(k^{(2)}_s,(k^{(2)}_s)^{-1})
    =\sum_{s=1}^{2t}\ln\kappa_s\),
    we obtain
    \begin{equation}
        \ln \pi_{2t}
        \;\approx\;
        \sum_{s=1}^{2t}\ln|\rho_s|
        \;+\;
        \sum_{s=1}^{t}\ln\bigl|\sin(\theta_{2s}-\theta_{2s-1})\bigr|
        \;+\;
        \sum_{s=1}^{2t}\ln\kappa_s~.
    \label{62th5}
    \end{equation}

    By the Central Limit Theorem, assuming all variances exist, this behaves as
    \begin{equation}
        \ln\pi_{2t} \sim
        \mathcal{N}\!\Big(2t(\ln\rho + \ln\kappa - \mu_\theta),\;
                        2t(\sigma_\rho^2 + \sigma_\kappa^2 + \sigma_\theta^2)\Big),
    \end{equation}
    with
    \begin{equation}
        \sum_{s=1}^t \ln|\sin(\theta_{2s}-\theta_{2s-1})|
        \sim \mathcal{N}(-2t\mu_\theta,\,t\sigma_\theta^2).
    \end{equation}

    We thus obtain the following approximations for the exponents:
    \begin{equation}
        \gamma \approx \ln\rho + \ln\kappa - \mu_\theta,
        \qquad
        \alpha \approx -\frac{2\gamma}{\sigma_\rho^2 + \sigma_\kappa^2 + \sigma_\theta^2}.
        \label{egyhyww}
    \end{equation}
    This shows explicitly how random rotations can strongly amplify growth:  
    if $\mu_\theta = \sigma_\theta = 0$, then at each step the rotation reinjects the system 
    perfectly along the most expanding direction (corresponding to 
    $|\sin(\theta_{2s}-\theta_{2s-1})|=1$).  
    In this idealized situation, the Lyapunov exponent nearly attains its upper bound 
    and the tail exponent its lower bound.  
    In practice, stochastic fluctuations in $\theta_t$ reduce this effect, 
    but the mechanism highlights how non-normality can push the system 
    much closer to instability even when it is not spectrally critical.

    \subsection{Numerical Application}

\begin{figure*}
        \centering
        \includegraphics[width=\textwidth]{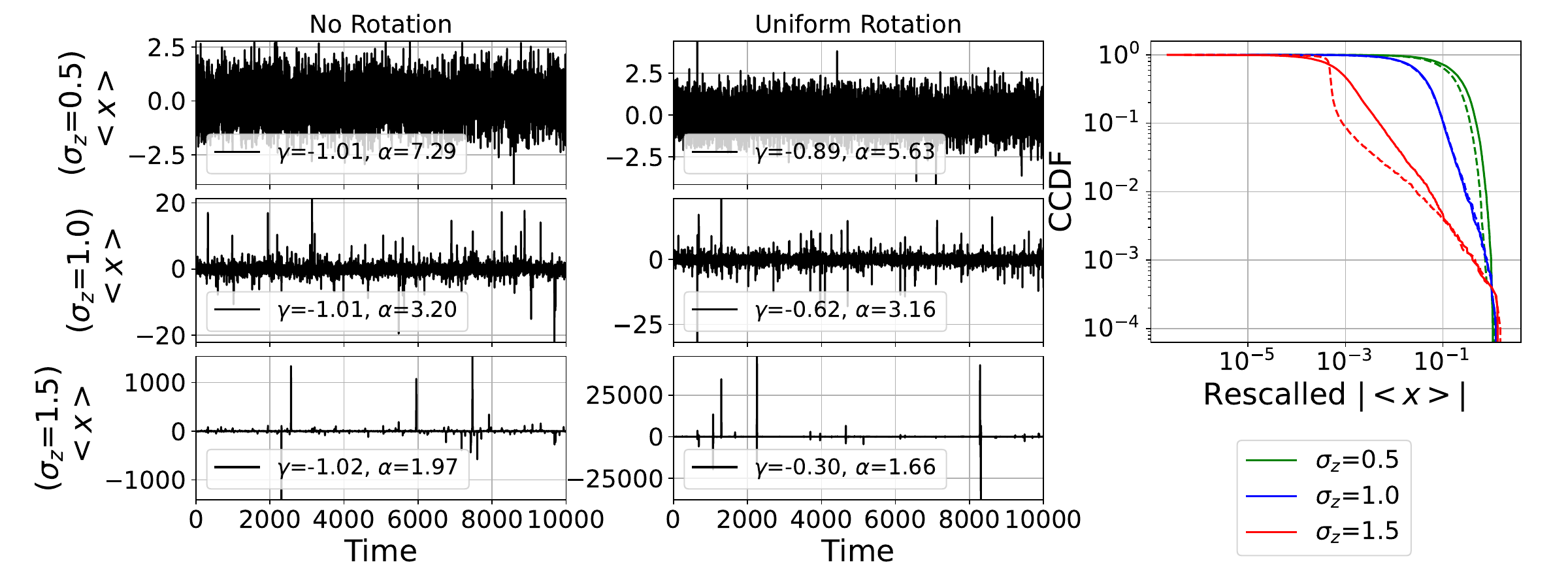}
        \caption{
            Simulation of the two-dimensional Kesten process \eqref{eq:apx_kesten}, 
            with matrices $\A_t$ defined by \eqref{eq:apx_kesten_2d_num}.  
            Parameters are set to $\delta = 0$ and $\ln\rho = -1$ (hence $\sigma_\rho = 0$).  
            Left column: no rotation ($\theta_t = 0$).  
            Middle column: uniform rotation ($\theta_t \sim \mathcal{U}(0,2\pi)$).  
            Each row corresponds to a different level of non-normal variability, 
            from top to bottom: $\sigma_z = 0.5, 1, 1.5$.  
            The $y$-axis reports the crossing mean $\langle x\rangle$
            between the two components 
            of the system.  
            The right panel shows the complementary cumulative distribution function (CCDF) 
            of the absolute mean, normalized so that the fifth-largest observation equals~1.  
            Solid lines correspond to the no-rotation case, while dashed lines correspond 
            to uniform rotations.  
            The numerical procedures used to estimate the Lyapunov and tail exponents 
            are detailed in Appendix~\ref{apx:tools}.
        }
        \label{fig:apx_simulation_2d}
    \end{figure*}

    To illustrate the theoretical results derived in the previous sections, 
    we now turn to a numerical study of the two-dimensional Kesten process.  
    For this purpose, we use the generic random matrix (\ref{twrhy2bq})
    \begin{equation}    
    \label{eq:apx_kesten_2d_num}
        \A_t = \rho_t \U(\theta_t)
        \begin{pmatrix}
            \delta & (1-\delta)z_t \\
            (1-\delta)z_t^{-1} & \delta
        \end{pmatrix} \U(\theta_t)^\dag,
    \end{equation}
    where $\rho_t$ controls the spectral radius, $z_t$ characterizes non-normal variations, 
    $\delta$ quantifies spectral imbalance, and $\U(\theta_t)$ is a rotation matrix.  
    Throughout this section we assume that $\delta$ is constant,  
    $\rho_t$ and $z_t$ are log-normally distributed 
    ($\ln \rho_t \sim \mathcal N(\rho,\sigma_\rho^2)$ and
    $\ln z_t \sim \mathcal N(0,\sigma_z^2)$), and $\theta_t$ is either uniformly 
    distributed on $[0,2\pi]$ or fixed to zero.
    \newline
    
     Let us define the (log-)condition number by
    \(
    \ln \kappa_t := \bigl|\ln z_t\bigr|
    \).
    Let $X:=\ln z_t \sim \mathcal N(0,\sigma_z^2)$ and $Y:=|X|$ (a half-normal variable).  
    The pdf of $Y$ is
    \begin{equation}
        f_Y(y)=\frac{2}{\sqrt{2\pi}\,\sigma_z}\,e^{-y^2/(2\sigma_z^2)},\qquad y\ge 0.
    \end{equation}

    \begin{itemize}[label={}, leftmargin=0pt]
        \item \emph{Mean.}
        Using $E[Y]=\int_0^\infty y f_Y(y)\,dy$,
        \begin{equation}
            \begin{split}
                E[Y]
                &=\frac{2}{\sqrt{2\pi}\,\sigma_z}\int_0^\infty y\,e^{-y^2/(2\sigma_z^2)}\,dy \\
                &=\frac{2}{\sqrt{2\pi}\,\sigma_z}\,\Bigl[-\sigma_z^2 e^{-y^2/(2\sigma_z^2)}\Bigr]_{0}^{\infty} \\
                &=\sigma_z\sqrt{\frac{2}{\pi}}.
            \end{split}
        \end{equation}

        \item \emph{Second moment.} Since $Y^2=X^2$ and $E[X^2]=\sigma_z^2$ for a centered normal,
        \(
            E[Y^2]=E[X^2]=\sigma_z^2
        \).

        \item \emph{Variance.} Hence
        \begin{equation}
            \operatorname{Var}(Y)=E[Y^2]-\bigl(E[Y]\bigr)^2
            =\sigma_z^2-\sigma_z^2\frac{2}{\pi}
            =\Bigl(1-\frac{2}{\pi}\Bigr)\sigma_z^2.
        \end{equation}
    \end{itemize}

    Identifying $Y=\ln\kappa_t$, we obtain
    \begin{equation}
        \begin{split}
            &\ln\kappa \;=\; E[\ln\kappa_t]=\sigma_z\sqrt{\frac{2}{\pi}}, \\
            &\sigma_\kappa^2 \;=\; \operatorname{Var}(\ln\kappa_t)
            =\Bigl(1-\frac{2}{\pi}\Bigr)\sigma_z^2~.
        \end{split}
        \label{7tbg10}
    \end{equation}

    For uniformly distributed rotations, $\theta_t\sim \mathcal{U}(0,2\pi)$, 
    the differences $\theta_t - \theta_{t+1}$ are also i.i.d.~uniform on $[0,2\pi]$.  
    This implies
    \begin{equation}
        \mu_\theta = \ln 2,
        \qquad
        \sigma_\theta^2 = \frac{\pi^2}{12}.
    \end{equation}
    Hence, the system is governed by four hyper-parameters: 
    $\rho$, $\sigma_\rho$, $\sigma_z$, and $\delta$.
    \newline

  \begin{figure*}
        \centering
        \includegraphics[width=\textwidth]{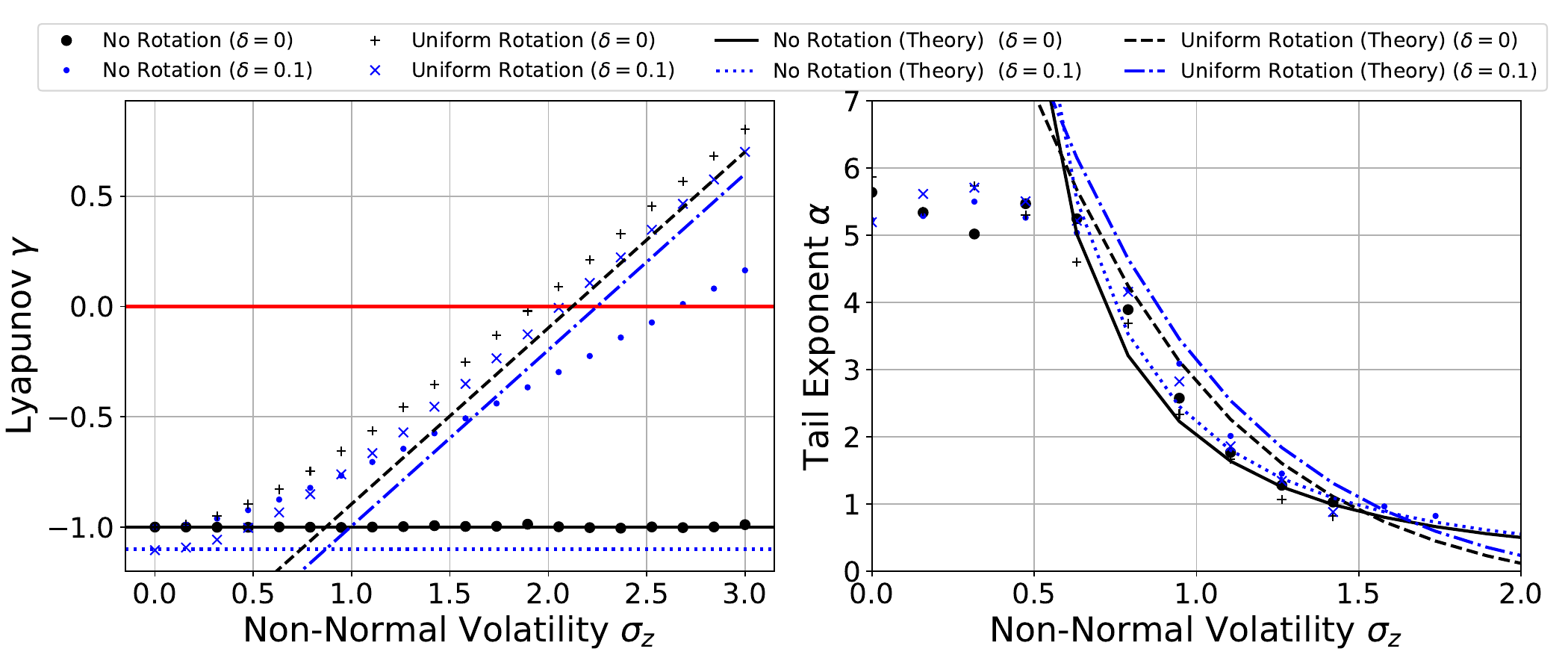}
        \includegraphics[width=\textwidth]{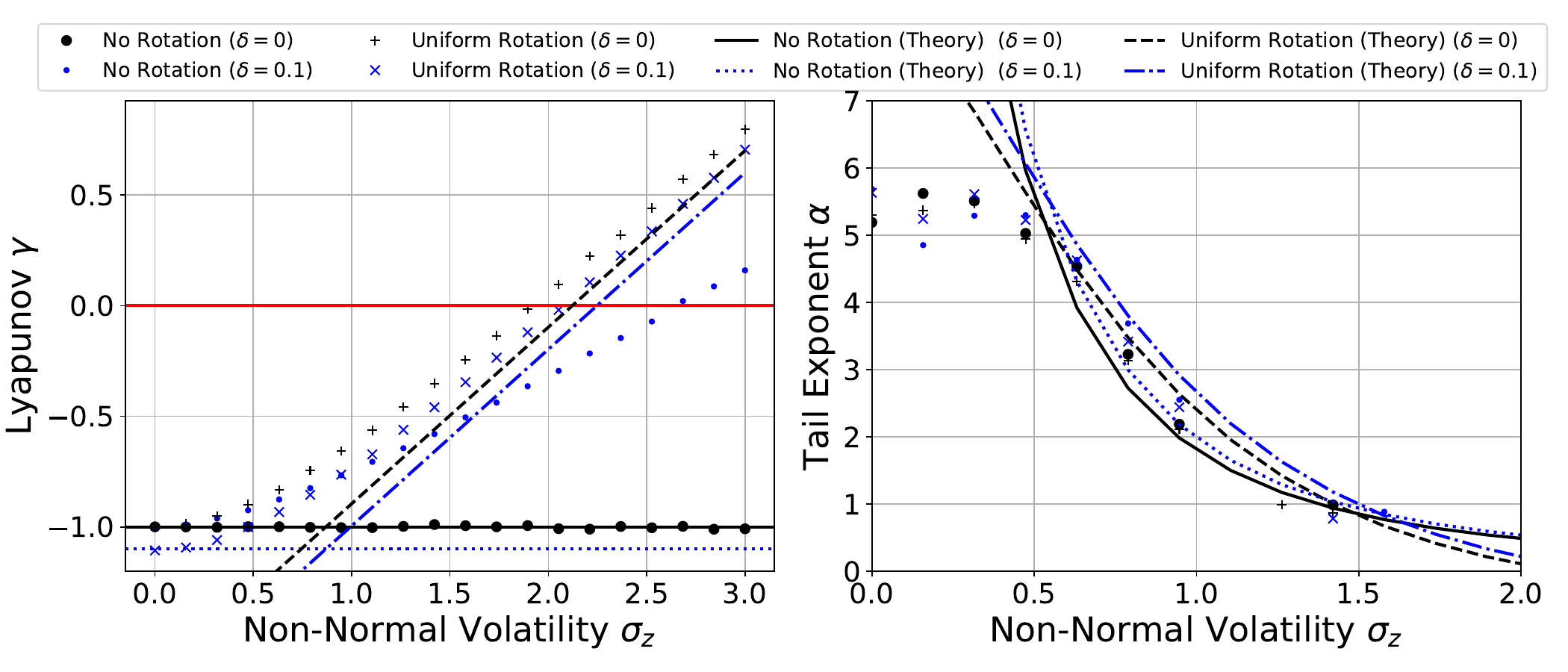}
        \caption{
            Study of the two-dimensional Kesten process \eqref{eq:apx_kesten}, 
            with matrices $\A_t$ defined by \eqref{eq:apx_kesten_2d_num}.  
            The system is simulated with $\ln\rho_t=-1$ (top) and $\ln\rho_t \sim \mathcal{N}(-1,1/9)$ (bottom), 
            while varying the non-normal variance $\sigma_z$ and the imbalance parameter $\delta$, 
            both with uniform rotations ($+$) and without rotations ($\cdot$).  
            Theoretical predictions with rotations (dashed lines) and without rotations 
            (solid lines) are taken from \eqref{eq:apx_case_2} and \eqref{eq:apx_case_1}, respectively.
            We considered the case without any imbalance i.e. $\delta=0$; (black)
            and with imbalance i.e. $\delta=0.1$; (blue).
            Left panel: Lyapunov exponent $\gamma$ as a function of $\sigma_z$.
            Right panel: corresponding tail exponent $\alpha$ (see Section~\ref{apx:tools} 
            for details of the numerical estimation).
            We zoomed the results on $\alpha\in[0,7]$, where the missing points are due to a diverging measure of the tail exponent,
            when it tends to be close to zeros and/or the Lyapunov exponent becomes positive.
        }
        \label{fig:apx_lyapunov_2d}
    \end{figure*}

    We focus on two limiting scenarios:
    \begin{itemize}
        \item \emph{Case 1: No Rotation.} $\theta_t = 0$.  
        In this case the theoretical predictions reduce to
        \begin{equation}    \label{eq:apx_case_1}
            \begin{split}
                &\gamma = \ln\rho - \delta + \mathcal{O}(\delta^2), \\
                &\alpha = -\frac{2\gamma}{\sigma_\rho^2 + \sigma_z^2} + \mathcal{O}(\delta^2).
            \end{split}
        \end{equation}
        \item \emph{Case 2: Uniform Rotation.} $\theta_t \sim \mathcal{U}(0,2\pi)$.  
        Here, the Lyapunov and tail exponents become
        \begin{equation}    \label{eq:apx_case_2}
            \begin{split}
                &\gamma \approx \ln\rho + \sqrt{\tfrac{2}{\pi}}\,\sigma_z - \ln 2 - \delta + \mathcal{O}(\delta^2), \\
                &\alpha \approx -\frac{2\gamma}{\sigma_\rho^2 + \big(1-\tfrac{2}{\pi}\big)\sigma_z^2 + \tfrac{\pi^2}{12}} + \mathcal{O}(\delta^2).
            \end{split}
        \end{equation}
    \end{itemize}

    Figure~\ref{fig:apx_simulation_2d} shows simulations for the case $\delta=0$ 
    with $\ln\rho_t=-1$ fixed, and for increasing non-normal variability 
    $\sigma_z = 0.5,\,1,\,1.5$.  We measure both 
    the Lyapunov exponent $\gamma$ and the tail exponent $\alpha$ 
    (see Appendix~\ref{apx:tools} for methodology).  
    We compare the no-rotation and uniform-rotation cases and show in the first two columns
   the temporal evolution of the crossing mean $\langle x\rangle$ between the two components 
    of the system. The crossing mean $\langle x\rangle$ is defined as follows.
   Let $x_t = (x^{(1)}_t, x^{(2)}_t)^\top$ denote the two components of the 
    two-dimensional Kesten process. Since the dynamics alternates between the two
    coordinates, a convenient scalar observable is obtained by averaging them at
    their crossings. We define
    \[
    \langle x_t \rangle \;=\; \frac{1}{2}\Bigl(x^{(1)}_t + x^{(2)}_t\Bigr),
    \]
    measured at the instants $t$ where $x^{(1)}_t$ and $x^{(2)}_t$ cross each
    other (i.e. when $x^{(1)}_t = x^{(2)}_t$ or $x^{(1)}_t = -\,x^{(2)}_t$).
    This ``crossing mean'' provides a single representative trajectory that
    summarizes the joint growth of the two components.
    
    For \emph{no rotation}, increasing $\sigma_z$ decreases the tail exponent 
    while leaving the Lyapunov exponent near $\gamma=-1$, as predicted (\ref{eq:apx_case_1}).  
    For \emph{uniform rotation}, $\gamma$ increases while $\alpha$ decrease with $\sigma_z$ (\ref{eq:apx_case_2}).  
    In particular, $\gamma$ vanishes at a critical value of $\sigma_z$ given by $\sigma_c \approx 2.1$, 
    indicating the onset of instability.
   
    Figure~\ref{fig:apx_lyapunov_2d} further explores these trends by plotting 
    $\gamma$ and $\alpha$ as functions of $\sigma_z$ for the case of a constant spectral radius $\ln\rho_t=-1$,
    and $\ln\rho_t \sim \mathcal{N}(-1,1/9)$. 
    Here, the theoretical tail exponent for $\delta=0$, no rotation, and $\sigma_z=0$ 
    is $\alpha=-10$.  
    The numerical results confirm that, with no rotation, the Lyapunov exponent remains 
    constant, while with uniform rotation $\gamma$ grows approximately linearly with 
    $\sigma_z$, with slope $\sqrt{2/\pi}$ as predicted.  
    However, a finite-size “transition region” appears for small $\sigma_z$, where 
    the linear theory does not apply.

    For the tail exponent, the measured values follow the theoretical trend but do not 
    exactly match for small non-normality, where the measured $\alpha$ tends to saturate.
    This discrepancy is expected: estimating $\alpha$ from finite samples is notoriously 
    difficult, especially when $\alpha$ is large or close to unity, and may yield 
    spurious or divergent results.
    \newline

    Finally, we investigate the effect of nonzero $\delta$.  
    We observe that, with uniform rotation, the system behaves as predicted,
    with a shift given by $-\delta$,
    but in the absence of rotation,
    we also observed that the system exhibits a growth in the Lyapunov exponent due to the non-normal behavior,
    revealing an additional destabilizing mechanism occurring at higher order power in $\delta$.

    \subsection{Synthesis}

    In this section, we have explored the two-dimensional Kesten process as a tractable yet 
    sufficiently rich setting to understand the interplay between spectral properties, non-normality, and critical behavior.  
    Starting from an exactly solvable case,
    we showed that the Lyapunov exponent depends only on the spectral radius,
    while the tail exponent is rescaled by the variance of non-normal fluctuations.  
    Through perturbative expansions, we extended these results to strong non-normal matrices,
    and demonstrated that rotation dynamics can either suppress or reinforce instabilities by reinjecting trajectories into expanding directions.  
    Finally, numerical simulations confirmed the theoretical predictions,
    while also highlighting subtle finite-size effects and nonlinear corrections when spectral imbalance is present.

    The two-dimensional case thus provides a clear picture:
    non-normality does not necessarily alter the stability criterion (through the Lyapunov exponent)
    but it can significantly reduce the tail exponent,
    effectively pushing the system into heavy-tailed regimes.  
    When combined with stochastic rotations and/or spectral imbalance (via the parameter $\delta$),
    non-normality can even shift the system toward true criticality,
    despite being spectrally subcritical.

    In the next section, we move beyond the two-dimensional setting and investigate the generalization to \(N\)-dimensional Kesten processes.  
    This will allow us to understand how non-normality scales with dimension,
    how collective effects emerge, and how the interplay of Lyapunov and tail exponents shapes the onset of critical phenomena in high-dimensional stochastic systems.


    \section{Generalized $N$-Dimensional Case}


    We now extend the analysis to the case of large dimension $N \gg 1$.  
    The aim is to derive approximate expressions for the Lyapunov exponent and the tail exponent, 
    and to identify how non-normality interacts with dimensionality to push the system closer to criticality.  
    To do so, we build on the spectral and singular value decompositions of the random matrices $\A_t$ 
    and apply tools from random matrix theory and extreme value theory.
    \newline

    We begin with the decomposition
    \begin{equation}    \label{eq:apx_dec_n_dim}
        \A_t \;=\; \U_t \Sig_t \V_t^\dagger \,\Lamb_t\, \V_t \Sig_t^{-1} \U_t^\dagger~,
    \end{equation}
   where
    \begin{itemize}[label={}, leftmargin=0pt]
        \item $\Lamb_t = \text{Diag}(\lambda_{1,t},\dots,\lambda_{N,t})$ contains the eigenvalues,
        \item $\PP_t = \U_t \Sig_t \V_t^\dag$ is the eigenbasis transformation matrix,
        \item $\Sig_t=\mathrm{Diag}(s_{1,t},\dots,s_{N,t})$ the singular values of $\PP_t$,
        \item $\U_t$ and $\V_t$ are unitary matrices.
    \end{itemize}

    We assume that the eigenvalues $\{\lambda_{i,t}\}$, singular values $\{s_{i,t}\}$, 
    and the rotations $\U_t, \V_t$ are independent, and that all degrees of freedom are i.i.d. in time.

    \subsection{Dominant Approximation via Condition Number}

    Starting from the SVD-based decomposition (\ref{eq:apx_dec_n_dim}),  
    we write the singular-vector expansions
    \begin{equation}
    \Sig_t \;=\; \sum_{i=1}^N s_{i,t}\, \uu_{i,t}\vv_{i,t}^\dagger,
    \;
    \Sig_t^{-1} \;=\; \sum_{i=1}^N s_{j,t}^{-1}\, \vv_{i,t}\uu_{i,t}^\dagger,
    \end{equation}
    so that     
    \begin{equation}
        \begin{split}
            \Sig_t\, \V_t^\dagger \Lamb_t \V_t \,\Sig_t^{-1}
            &= \sum_{i,j=1}^N \frac{s_{i,t}}{s_{j,t}}
            \;\uu_{i,t}\Bigl(\vv_{i,t}^\dagger \V_t^\dagger \Lamb_t \V_t \vv_{j,t}\Bigr)\uu_{j,t}^\dagger \\
            &= \sum_{i,j=1}^N \frac{s_{i,t}}{s_{j,t}} \;\lambda_{ij,t}\; \uu_{i,t}\uu_{j,t}^\dagger,
        \end{split}
    \end{equation}
    where we denote
    \begin{equation}
    \lambda_{ij,t}\;:=\;\bigl(\V_t^\dagger \Lamb_t \V_t\bigr)_{ij}.
    \end{equation}
    Let $s_{\max,t}$ and $s_{\min,t}$ be the extreme (largest and smallest) singular values of $\Sig_t$, with
    corresponding singular vectors $\uu_{\max,t}$, $\uu_{\min,t}$. Then,  the condition number 
    $ \kappa_t = \|\PP_t\|\,\|\PP_t^{-1}\|$  (\ref{dhwgbwa}) is given by     
    $\kappa_t:=s_{\max,t}/s_{\min,t}$. Then in the double sum above,
    the coefficient $s_{i,t}/s_{j,t}$ is maximized at $(i,j)=(\max,\min)$ and equals $\kappa_t$,
    while for any other pair $(i,j)$ one has $s_{i,t}/s_{j,t}\le \kappa_t$ with strict inequality unless
    $(i,j)=(\max,\min)$. If we moreover assume independence and i.i.d.\ in time of
    $\{\lambda_{ij,t}\}$ and that these entries do not concentrate pathologically on subdominant $(i,j)$ pairs,
    the rank-one term with $(i,j)=(\max,\min)$ dominates:
    \begin{equation}
    \Sig_t\, \V_t^\dagger \Lambda_t \V_t \,\Sig_t^{-1}
    \;\approx\; \kappa_t \,\lambda_{\,\max,\min,t}\;\uu_{\max,t}\uu_{\min,t}^\dagger.
    \end{equation}
    Left- and right-multiplication by the unitary $\U_t$ leaves the operator norm unchanged, so inserting
    back into $\A_t$ yields the dominant-direction (rank-one) approximation
    \begin{equation}
        \A_t \;\approx\; \kappa_t\,\lambda_t\;\uu_{\max,t}\uu_{\min,t}^\dagger
        ,\;
        \lambda_t:=\bigl(\V_t\Lamb_t \V_t^\dagger\bigr)_{\max,\min}~.
    \label{thty276}
    \end{equation}
    The approximation (\ref{thty276}) is valid in the \emph{ill-conditioned, large-$N$ (rank-one-dominance)} regime:
    \begin{itemize}
        \item $\kappa_t\gg 1$ so that the ratio $s_{i,t}/s_{j,t}$ is sharply peaked at $(i,j)=(\max,\min)$;

        \item the spectral mixing $\V_t^\dagger\Lamb_t \V_t$ has entries $\lambda_{ij,t}$ with bounded moments and
        no anomalous amplification on subdominant pairs $(i,j)\neq(\max,\min)$;

        \item $\U_t,\V_t$ are (approximately) Haar-distributed and independent of $\Sig_t,\Lamb_t$,
        as assumed in the large-$N$ model setup.
    \end{itemize}
    Under these conditions (typical at large $N$, where extreme singular values separate),
    the outer-product term $\uu_{\max,t}\uu_{\min,t}^\dagger$
    provides the leading contribution to $\A_t$, and \eqref{thty276} controls the growth in the product $\PPi_t$ used later.

    Under this approximation, the product of matrices becomes
    \begin{equation}    \label{eq:apx_pi_nd}
        \PPi_t \approx \Bigg(\prod_{s=1}^t \kappa_s \lambda_s\Bigg)
        \Bigg(\prod_{s=1}^{t-1} \uu_{\min,s}\cdot\uu_{\max,s+1}\Bigg)
        \uu_{\max,1}\uu_{\min,t}^\dag,
    \end{equation}
    so that the product norm satisfies
    \begin{equation}
        \ln\pi_t \approx \sum_{s=1}^t \ln|\lambda_s|
        + \sum_{s=1}^t \ln\kappa_s
        + \sum_{s=1}^{t-1}\ln\big|\uu_{\min,s}\cdot\uu_{\max,s+1}\big|.
    \end{equation}

    By the central limit theorem, in the limit $t\to\infty$, the distribution of $\ln\pi_t$ is approximately Gaussian,
    \begin{equation}    \label{eq:apx_n_dim_prod}
        \begin{split}
            &\ln\pi_t \sim \mathcal{N}\!\Big(t(\ln\lambda + \ln\kappa + \ln\mu_\theta),
            \; t(\sigma_\lambda^2 + \sigma_\kappa^2 + \sigma_\theta^2)\Big), \\
            &\ln\rho := \mathbb{E}[\ln|\lambda_1|],\quad
            \sigma_\rho^2 := \text{Var}[\ln|\lambda_1|], \\
            &\ln\kappa := \mathbb{E}[\ln \kappa_1],
            \quad\sigma_\kappa^2 := \text{Var}[\ln \kappa_1], \\
            &\ln\mu_\theta := \mathbb{E}\!\left[\ln\big|\uu_{\min,1}\cdot\uu_{\max,2}\big|\right], \\
            &\sigma_\theta^2 := \text{Var}\!\left[\ln\big|\uu_{\min,1}\cdot\uu_{\max,2}\big|\right].
        \end{split}
    \end{equation}

    Thus, the Lyapunov exponent depends on three contributions: the eigenvalue statistics, the condition number of $\PP_t$, and the reinjection term due to random orientations.

    \subsection{Scaling of the Contribution with Dimension $N$}

    For each of the parameters in \eqref{eq:apx_n_dim_prod}, we can characterize how it scales with the dimension $N$ of the system
    by using the law of large numbers, the central limit theorem, and extreme value theory (EVT).
    We now discuss each contribution in turn.
    \newline

    Let us define the average spectral contribution
    \begin{equation}
        \ln\rho := \mathbb{E}\!\left[\ln|\lambda_{i,t}|\right],
    \end{equation}
    and decompose each eigenvalue as
    \begin{equation}
        \lambda_{i,t} = e^{\ln\rho}\,\lambda_{i,t}^0,
        \qquad
        \mathbb{E}[\ln|\lambda_{i,t}^0|] = 0.
        \label{fdhhnbgwq}
    \end{equation}
    Hence,
    \begin{equation}
        \ln|\lambda_t| = \ln\rho + \ln\Bigg|\sum_k \lambda_{k,t}^0 \vv_{ik}\vv_{jk}^*\Bigg|,
        \label{dth3ynb2q}
    \end{equation}
    where $(i,j)=(i_{\max},i_{\min})$ are the indices associated with the largest and smallest singular values.  
    To obtain (\ref{dth3ynb2q}), we have used $\lambda_t:=\bigl(\V_t\Lamb_t \V_t^\dagger\bigr)_{\max,\min}$ as defined in (\ref{thty276}).
    
    The coefficients $\vv_{ik}$ and $\vv_{jk}$ are entries of two independent rows.
    The vectors $\vv_{i}$ and $\vv_{j}$ can be treated as independent isotropic random vectors on the unit sphere,
    so their overlaps have mean zero and variance $\mathcal{O}(1/N)$.
    Because the $\lambda_{k,t}^0$'s are i.i.d. with finite variance,
    the central limit theorem implies that
    \begin{equation}
        \sum_k \lambda_{k,t}^0 \vv_{ik}\vv_{jk}^*
        \;\;\overset{d}{\longrightarrow}\;\;
        \mathcal{N}\!\Big(0,\frac{\sigma_0^2}{N}\Big)~,
        \qquad N\gg 1.
    \end{equation}
    where $\sigma_0$  is the standard deviation of the fluctuations of the normalized eigenvalues $\lambda^0_{i,t}$ introduced in (\ref{fdhhnbgwq}). 
    Intuitively, the random overlaps $\vv_{ik}\vv_{jk}^*$ behave like $O(1/\sqrt{N})$ Gaussian coefficients,
    so the whole sum is a Gaussian random variable of amplitude $\sim 1/\sqrt{N}$.

    If $X\sim \mathcal{N}(0,\sigma_0^2/N)$, then $|X|$ has a folded normal distribution.  
    The statistics of $\ln|X|$ are well-known:
    \begin{equation}
        \begin{split}
            &\mathbb{E}[\ln|X|] = \ln\sigma_0 - \tfrac{1}{2}\ln N - \frac{g+\ln 2}{2}, \\
            &\text{Var}[\ln|X|] = \frac{\pi^2}{8},
        \end{split}
    \end{equation}
    where $g$ is Euler's constant.
    This follows from the fact that $X/\sigma_0\sqrt{N}$ is standard normal,
    and the distribution of $\ln|Z|$ for $Z\sim\mathcal{N}(0,1)$ is explicitly known.

    Combining these results, we obtain
    \begin{equation}
        \begin{split}
            &\ln\lambda = \ln\rho + \ln\sigma_0 - \frac{g+\ln 2}{2} - \frac{1}{2}\ln N, \\
            &\sigma_\lambda^2 = \frac{\pi^2}{8}.
        \end{split}
    \end{equation}
    Thus, the eigenvalue contribution to the Lyapunov exponent decreases as $-(1/2)\ln N$ in high dimensions,
    with Gaussian fluctuations of universal variance $\pi^2/8$.
    \newline

    We now analyze how the condition number
    \begin{equation}
        \kappa_t = \frac{s_{\max,t}}{s_{\min,t}}
    \end{equation}
    scales with the dimension $N$.  
    Taking logarithms, we have
    \begin{equation}
        \ln \kappa_t = \max_i \ln s_{i,t} - \min_i \ln s_{i,t}.
    \end{equation}
    We assume that the singular values are log-normally distributed, i.e.
    \begin{equation}
        \ln s_{i,t} \sim \mathcal{N}(0,\sigma^2),
        \qquad i=1,\dots,N,
        \label{rhwryh2g}
    \end{equation}
    independently across $i$.  
    Hence, $\ln s_{i,t}$ are $N$ i.i.d. Gaussian random variables.
    From EVT, the maximum of $N$ i.i.d. $\mathcal{N}(0,\sigma^2)$ variables satisfies
    \begin{equation}
        \max_i \ln\sigma_{i,t} \approx \sigma \sqrt{2 \ln N}
        + \frac{\sigma}{\sqrt{2\ln N}}\, G^+,
    \end{equation}
    where $G^+$ converges in distribution to a standard Gumbel variable.  
    Similarly, the minimum satisfies
    \begin{equation}
        \min_i \ln\sigma_{i,t} \approx -\sigma \sqrt{2 \ln N}
        + \frac{\sigma}{\sqrt{2\ln N}}\, G^-,
    \end{equation}
    with $G^-$ an independent Gumbel variable.
    Taking the difference, we obtain
    \begin{equation}
        \begin{split}
            \ln \kappa_t
            &= \max_i \ln s_{i,t} - \min_i \ln s_{i,t} \\
            &\approx 2\sigma \sqrt{2\ln N}
            + \frac{\sigma}{\sqrt{2\ln N}}\,(G^+ - G^-).
        \end{split}
    \end{equation}
    Thus the leading-order scaling of $\ln\kappa_t$ grows like $\sqrt{\ln N}$.
    The large $N$ dependence of $\ln\kappa_t$ is therefore
    \begin{equation}
        \ln \kappa \;\approx\; 2\sigma \sqrt{2\ln N},
    \end{equation}
    while the variance is controlled by the fluctuations of the Gumbel distribution.  
    Since $\mathrm{Var}(G^+ - G^-) = \pi^2/3$, we find
    \begin{equation}
        \sigma_\kappa^2 \;\approx\; \frac{\pi^2\sigma^2}{6\ln N}.
    \end{equation}
    This shows that the condition number diverges slowly as the dimension $N$ increases:
    the mean of $\ln\kappa_t$ grows like $\sqrt{\ln N}$, while its fluctuations shrink like $1/\sqrt{\ln N}$.  
    Hence, for large $N$, the instability induced by non-normality (measured through $\kappa_t$)
    becomes progressively more pronounced but also more concentrated around its typical value.
    \newline

    The last contribution in \eqref{eq:apx_n_dim_prod} comes from the scalar products 
    $\uu_{\min,t}\cdot\uu_{\max,t+1}$ between eigenvectors associated with the most contracting and expanding directions at successive times.  
    Assuming isotropy of the eigenbasis,
    both $\uu_{\max,t}$ and $\uu_{\min,t}$ can be treated as independent and uniformly distributed on the unit sphere in $\mathbb{R}^N$.

    For two independent isotropic unit vectors in $\mathbb{R}^N$, the dot product satisfies
    \begin{equation}
        \sqrt{N}\,\uu_{\min,t}\cdot\uu_{\max,t} \;\overset{d}{\longrightarrow}\; u_t \sim \mathcal{N}(0,1),
        \qquad N\gg 1.
    \end{equation}
    Thus, the overlap is typically of order $1/\sqrt{N}$.
    Taking logarithms, we obtain
    \begin{equation}
        \ln|\uu_{\min,t}\cdot\uu_{\max,t}|
        = -\tfrac{1}{2}\ln N + \ln|u_t|,
    \end{equation}
    where $u_t$ is standard Gaussian.  
    Therefore, the reinjection term is a deterministic shift $-\tfrac{1}{2}\ln N$ plus Gaussian-logarithmic fluctuations,
    for which it is well known that for $u_t\sim\mathcal{N}(0,1)$,
    \begin{equation}
        \mathbb{E}[\ln|u_t|] = -\frac{g+\ln 2}{2},
        \qquad
        \text{Var}[\ln|u_t|] = \frac{\pi^2}{8},
    \end{equation}
    where $g$ is Euler's constant.
    Combining these contributions, the reinjection term has
    \begin{equation}
        \ln\mu_\theta = -\tfrac{1}{2}\ln N - \tfrac{g+\ln 2}{2},
        \qquad
        \sigma_\theta^2 = \frac{\pi^2}{8}.
        \label{ryjtun3thw}
    \end{equation}
    The overlap between expanding and contracting directions shrinks like $N^{-1/2}$,
    which produces a negative correction to the Lyapunov exponent.  
    At the same time, the variance remains constant ($\pi^2/8$),
    reflecting universal logarithmic fluctuations associated with Gaussian overlaps.  
    Hence, reinjection acts as a dimensional penalty on stability,
    counterbalancing the growth induced by the condition number.

    \subsection{Asymptotic Lyapunov and Tail Exponents}

    Putting everything together, the Lyapunov exponent is approximated as
    \begin{equation}    \label{eq:apx_lyapunov_nd}
        \gamma \approx \ln\rho + \ln\sigma_0 - \ln N - \ln 2 - g + 2\sigma\sqrt{2\ln N}.
    \end{equation}
    Hence, for large $N$, there exists a critical non-normal standard deviation $\sigma_c(N)$ 
    of the singular values defined in (\ref {rhwryh2g})
    such that $\gamma \approx 0$:
    \begin{equation}    \label{eq:apx_sig_c_N}
        \sigma_c(N) = \frac{1}{2\sqrt{2}}\,\sqrt{\ln N}
        \left[1 + \frac{\ln 2 + g - \ln\rho - \ln\sigma_0}{\ln N}\right].
    \end{equation}
    
    From the CLT decomposition $\alpha \approx -\frac{2\gamma}{\sigma_\rho^2 + \sigma_\kappa^2 + \sigma_\theta^2}$
    derived in (\ref{egyhyww}), the variance entering the denominator of (\ref{egyhyww}) is the sum of the
    three contributions:  
    \[
    \sigma_\rho^2+\sigma_\kappa^2+\sigma_\theta^2
    \;\approx\; \underbrace{\sigma_\rho^2}_{\text{fixed}}
    \;+\;\underbrace{\Big(3+\frac{2\sigma^2}{\ln N}\Big)}_{\text{EVT range of Gaussians}}
    \;+\;\underbrace{\frac{\pi^2}{8}}_{\text{from (\ref{ryjtun3thw})}}.  
    \]
    Collecting constants and expanding consistently to first order around $\sigma=\sigma_c(N)$
    leads to the compact near-critical form 
    \begin{equation}
    \alpha \;\approx\;
    \frac{\sqrt{2\ln N}}{\;\dfrac{48}{\pi^2}\;}\;
    \frac{\;\sigma_c(N)-\sigma\;}{\;3+\dfrac{2\sigma^2}{\ln N}\;}
    \;+\; o\!\Big(\frac{1}{\sqrt{\ln N}}\Big)~,
    \label{1rth2ty01}
    \end{equation}
    Thus, the tail exponent decreases \emph{linearly} with the distance to criticality, with a slope that grows
    like $\sqrt{2\ln N}$ and a denominator collecting the variance contributions of spectrum, condition number,
    and reinjection. 
    These results show that, in high dimensions, the combined effect of spectral fluctuations, condition numbers, and random reinjection 
    induces a delicate balance between stability and criticality.  
    In particular:
    \begin{itemize}
        \item The critical non-normal variance $\sigma_c(N)$ grows like $\sqrt{\ln N}$, 
        implying that larger systems tolerate more heterogeneity before destabilizing when $\gamma\approx 0$.
        \item The tail exponent $\alpha$ decreases as $\sigma$ approaches $\sigma_c(N)$, 
        confirming that high-dimensional non-normality drives heavy-tailed fluctuations 
        even when the system is spectrally stable.
    \end{itemize}
    This highlights how non-normal amplification, coupled with dimensionality, 
    pushes the system closer to criticality in the sense of diverging moments and vanishing tail exponents.
    
    \subsection{Singular Values Random Perturbation}

    In the previous section, we focused on the case $\mathbb{E}[\ln\kappa_1]\gg 1$, which allowed us to approximate the Lyapunov exponent using the dominant singular value (condition number) only.  
    Here, we consider the opposite regime, where the deviations from normality are small and can be treated perturbatively.  
    To this end, we write
    \begin{equation}
        \ln s_{i,t} = \sigma \epsilon_{i,t}, \quad 
        \mathbb{E}[\epsilon_{i,t}] = 0, \quad 
        \text{Var}[\epsilon_{i,t}] = 1,
    \end{equation}
    with $\sigma \ll 1$ controlling the amplitude of the non-normal perturbation.

    Using the matrix decomposition \eqref{eq:apx_dec_n_dim}, we can expand
    \begin{equation}
        \begin{split}
            \A_t &= \A_t^0 + \sigma \A_t^1 + \mathcal{O}(\sigma^2),\\
            \text{where}\quad 
            \A_t^0 &= \U_t \V_t^\dag \Lamb_t \V_t \U_t^\dag,\\
            \A_t^1 &= \U_t \EE_t \V_t^\dag \Lamb_t \V_t \U_t^\dag - \U_t \V_t^\dag \Lamb_t \EE_t \V_t \U_t^\dag,
        \end{split}
    \end{equation}
    and $\EE_t = \mathrm{Diag}(\epsilon_{1,t},\dots,\epsilon_{N,t})$,
    so that $\A_t^0$ is Hermitian (symmetric) and $\A_t^1$ is anti-Hermitian (anti-symmetric).  
    This allows us to expand the product
    \begin{equation}
        \begin{split}
            \PPi_t &= \PPi_t^0 + \sigma \PPi_t^1 + \mathcal{O}(\sigma^2),\\
            \text{where}\quad
            \PPi_t^0 &= \prod_{s=1}^t \A_s^0,\\
            \PPi_t^1 &= \sum_{s=1}^t \left(\prod_{u=1}^{s-1}\A_u^0\right)\A_s^1 \left(\prod_{u=s+1}^{t}\A_u^0\right).
        \end{split}
        \label{dhtytqgb}
    \end{equation}

    Since $\PPi_t^0$ is Hermitian and $\sigma \ll 1$, the $L_2$-norm of the product can be expanded via standard eigenvalue perturbation theory as
    \begin{equation}
        \begin{split}
            \|\PPi_t\| \approx \|\PPi_t^0\|\Bigg[ &1 + \sigma \frac{\vv_{\max}^\dag \PPi_t^1 \vv_{\max}}{\|\PPi_t^0\|}  \\
            &\;+ \sigma^2 \frac{1}{\|\PPi_t^0\|} \sum_{k\neq \max} \frac{\left|\vv_k^\dag \PPi_t^1 \vv_{\max}\right|^2}{\|\PPi_t^0\| - \lambda_k} \Bigg],
        \end{split}
    \end{equation}
    where $\vv_{\max}$ is the eigenvector of $\PPi_t^0$ associated with the largest eigenvalue $\|\PPi_t^0\|$,
    and $\vv_k$ and $\lambda_k$ denote the other eigenvectors and eigenvalues of $\PPi_t^0$.  
    We note that the first-order correction $\PPi_t^1$ in the expansion (\ref{dhtytqgb})
contains exactly one occurrence of the perturbation $\A_t^s$, while all other factors are the unperturbed Hermitian matrices $\A_u^0$. Because $\A_t^s$ is anti-Hermitian, its eigenvalues are purely imaginary, so it generates infinitesimal unitary rotations. This can be seen by writing, for any vector $v$,
\begin{equation}
\frac{d}{d\epsilon}\,\| (I+\epsilon \A_t^s) v\|^2\Big|_{\epsilon=0}
= 2\,v^\dagger \tfrac{1}{2}(\A_t^s + \A_t^{s\dagger})v = 0,
\end{equation}
since $\A_t^s + \A_t^{s\dagger} = 0$. Therefore, the action of $\A_t^s$ changes only the \emph{direction} of $v$, not its norm, at first order in $\epsilon$. In other words, $\A_t^s$ acts as an infinitesimal generator of rotations in the complex vector space.

Because $\A_t^s$ has zero mean, $\E[\A_t^s] = 0$, the first-order correction $\PPi_t^1$ also averages to zero over the ensemble of random realizations. The reason is that each $\PPi_t^1$ contains a single insertion of $\A_t^s$, which upon averaging gives
\begin{equation}
\E[\PPi_t^1] = \sum_{s=1}^t 
\E\!\left[
\left(\prod_{u=1}^{s-1} \A_u^0 \right)
\A_s^1
\left(\prod_{u=s+1}^{t} \A_u^0 \right)
\right] = 0,
\end{equation}
since $\E[\A_s^1] = 0$ by construction. Thus, any contribution to the norm at order $\sigma$ cancels on average.

Geometrically, the anti-Hermitian perturbation introduces only infinitesimal rotations of the eigenbasis of $\PPi_t^0$, which conserve the L$_2$ norm. Hence, the norm $\|\PPi_t\|$ is invariant at first order, and corrections to the Lyapunov exponent arise only at order $\sigma^2$.  The physical interpretation is that anti-Hermitian fluctuations produce random rotational flows in matrix space that redistribute amplitudes among directions without altering the total norm to first order, so that only quadratic (variance-driven) effects contribute to the long-term growth rate.
    
    Moreover, since $\PPi_t^1$ is a sum of random matrices with finite variance,
    by the Central Limit Theorem, its matrix elements tend to a Gaussian distribution with zero mean for large $t$.  
    Consequently, the logarithm of the norm can be approximated as
    \begin{equation}
        \ln \pi_t \sim \mathcal{N}\Big(t (\gamma_0 + \sigma^2 c_1),\ t (\sigma_0^2 + c_2 \sigma^2)\Big), \quad t\to\infty,
    \end{equation}
    where $\gamma_0$ is the Lyapunov exponent of the purely normal system ($\sigma = 0$), $\sigma_0^2 t$ is the variance of $\ln \pi_t$ in the normal system, and $c_1,c_2 \ge 0$ are constants independent of $\sigma$ determined by the variance of the perturbation.

    Hence, for weak non-normality, the Lyapunov exponent and the associated tail exponent $\alpha$ admit the expansions
    \begin{equation}
        \gamma \approx \gamma_0 + c_1 \sigma^2, 
        \quad
        \alpha \approx -2 \frac{\gamma}{\sigma_0^2 + c_2 \sigma^2}.
    \end{equation}
    We see explicitly that, in this regime, the correction to the Lyapunov exponent is quadratic in $\sigma$,
    rather than linear, in agreement with the intuition that anti-Hermitian perturbations contribute only at second order to the growth rate.
   
    \subsection{Numerical application}
    
      \begin{figure*}
        \centering
        \includegraphics[width=\textwidth]{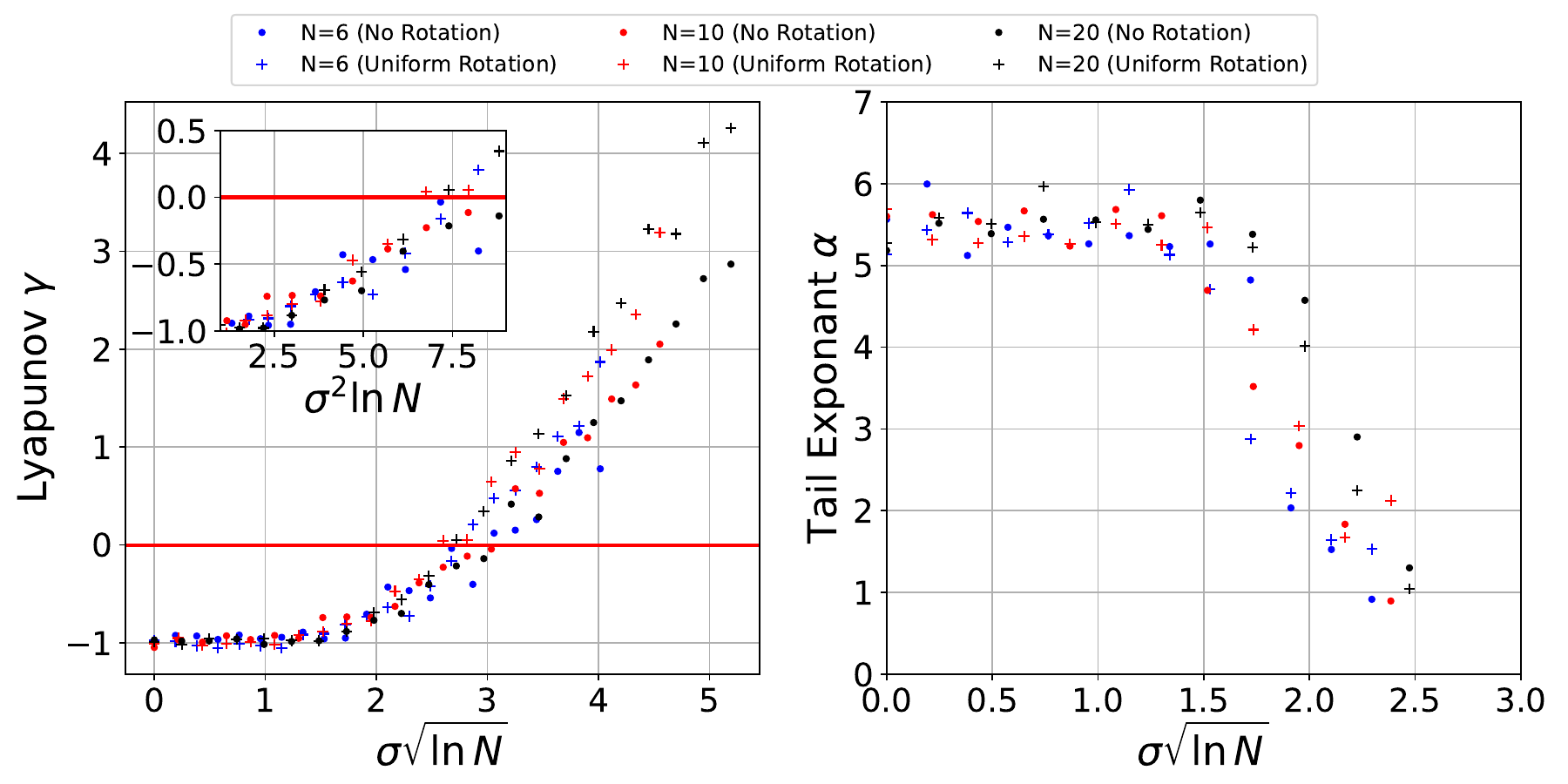}
        \includegraphics[width=\textwidth]{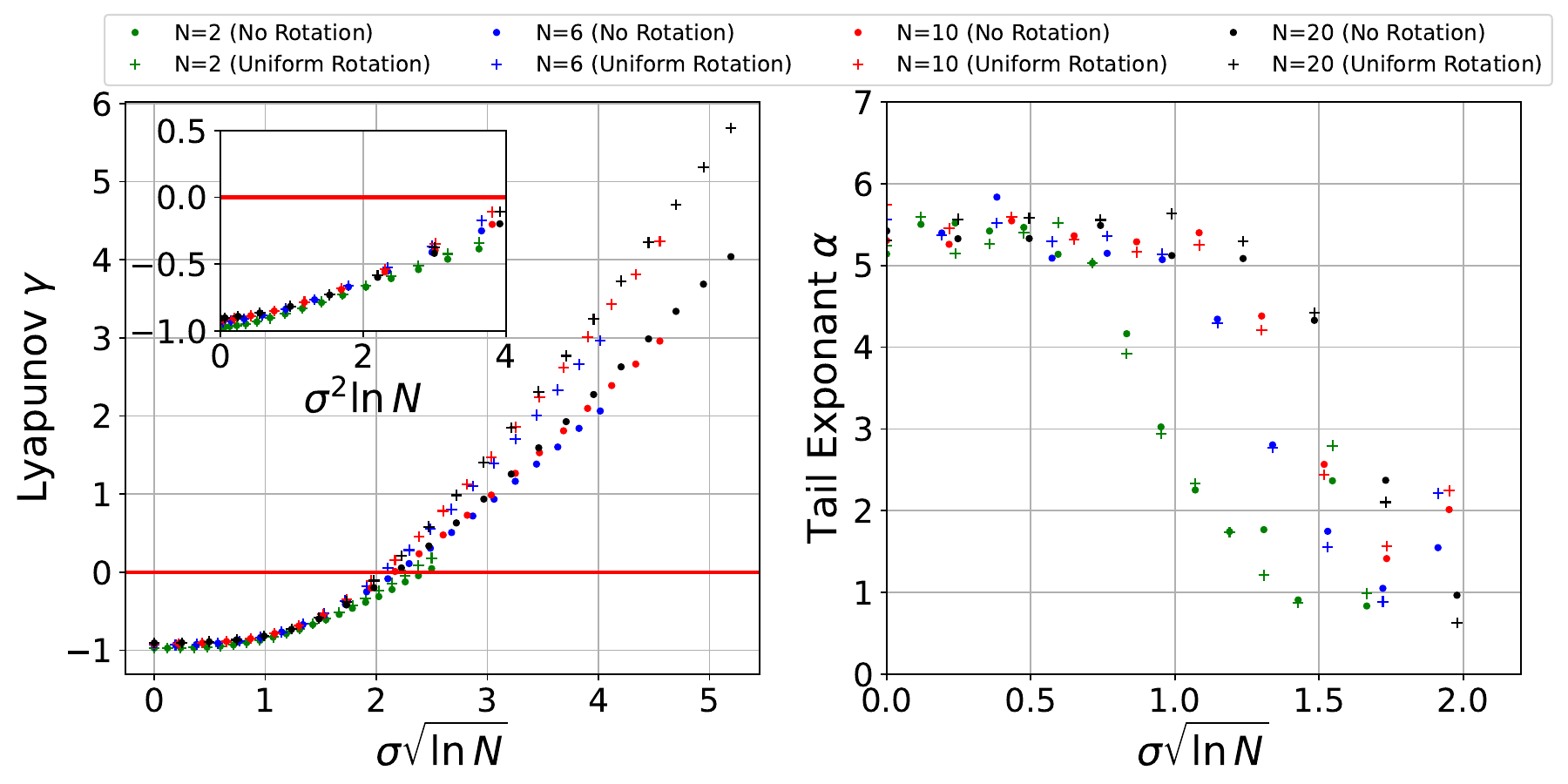}
        \caption{
            Lyapunov exponent \(\gamma\) for the Kesten process \eqref{eq:apx_kesten} with matrices \(\A_t\)
            generated via the decomposition \eqref{eq:apx_dec_n_dim} as a function of $\sigma \sqrt{\ln N}$, testing the scaling predictions.
            For the top panel, we generated $\V_t^\dag \Lambda_t \V_t$ once at the beginning and kept it fixed throughout the simulation,
            assuring that the ``normal'' part of the matrix is fixed.
            In the bottom panel we used \eqref{eq:apx_dec_n_dim} and generated the matrix $\V_t^\dag\Lamb_t\V_t$ at each step.
            We set
            \(\ln\lambda_{i,t}\overset{\text{i.i.d.}}{\sim}\mathcal{N}(-1,1/9)\) and
            \(\ln s_{i,t}\overset{\text{i.i.d.}}{\sim}\mathcal{N}(0,\sigma^2)\). Each curve shows
            \(\gamma\) as a function of \(\sigma\) for different dimensions $N$.
            The ``No rotation'' fixes \(\U_t=\I\); the ``Uniform rotation''
            samples \(\U_t\) uniformly at each time. Results are obtained using the QR-based
            Lyapunov estimator described in Appendix~\ref{apx:tools}.
        }
        \label{fig:apx_lyapunov_nd}
    \end{figure*}

    \begin{figure*}
        \centering
        \includegraphics[width=\textwidth]{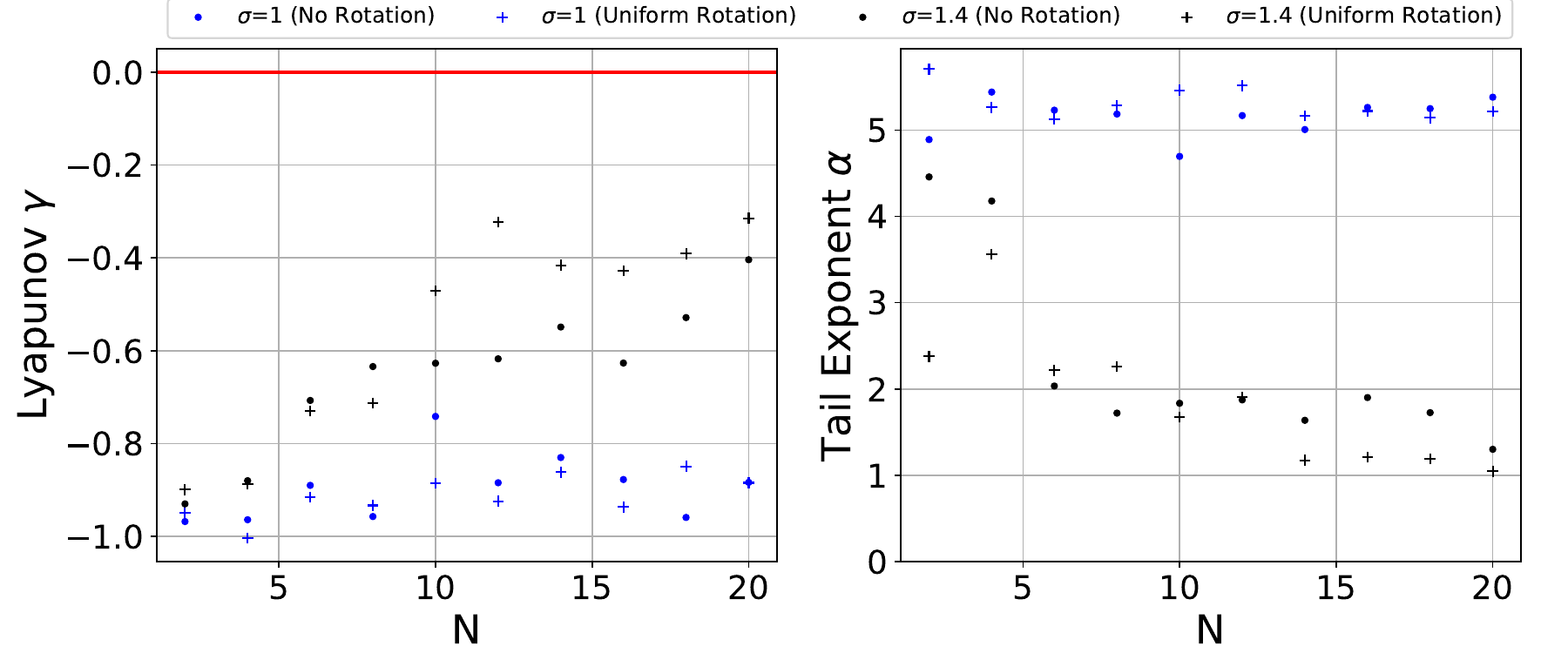}
        \includegraphics[width=\textwidth]{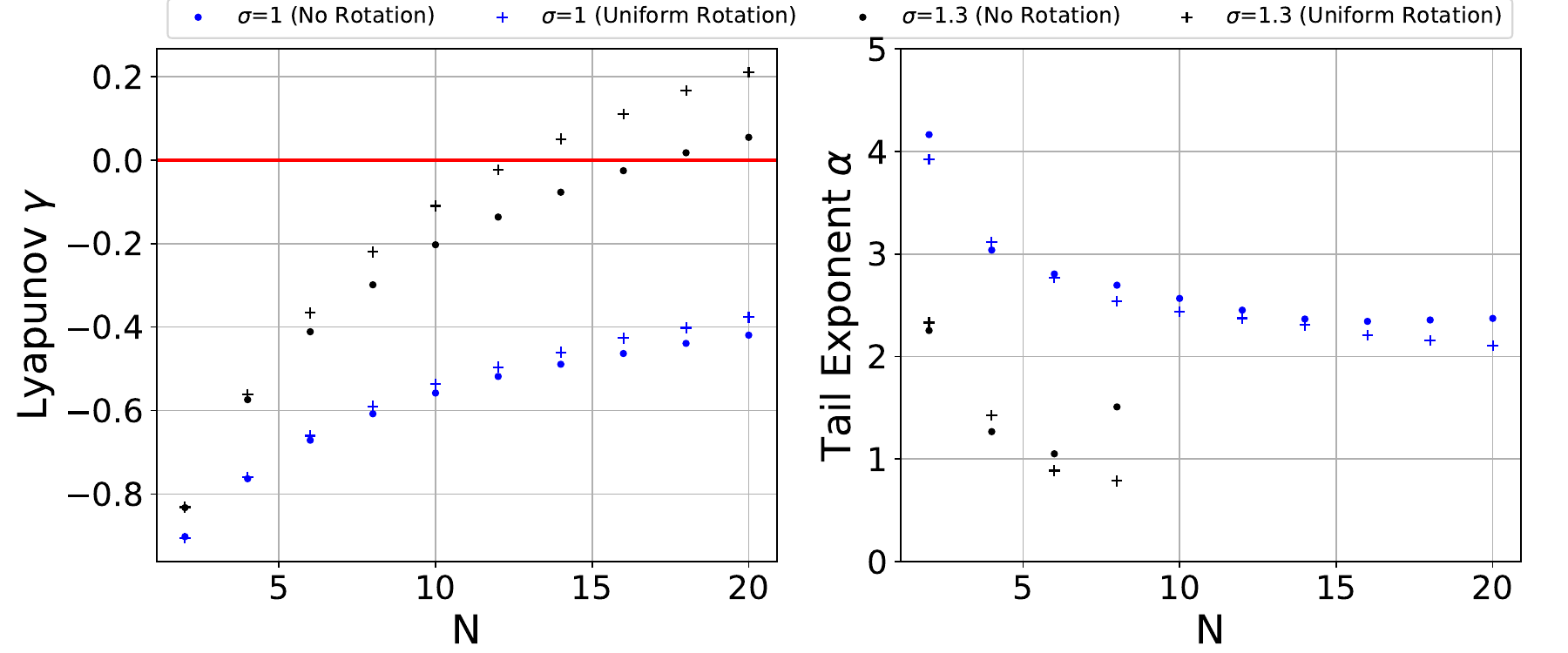}
        \caption{
            Lyapunov exponent \(\gamma\) for the Kesten process \eqref{eq:apx_kesten} with matrices \(\A_t\)
            generated via the decomposition \eqref{eq:apx_dec_n_dim}, as a function of the system dimension $N$.
            For the top panel, we generated $\V_t^\dag \Lambda_t \V_t$ once at the beginning and kept it fixed throughout the simulation, 
            ensuring that the ``normal'' part of the matrix remained constant.
            In the bottom panel, we used \eqref{eq:apx_dec_n_dim} and generated the matrix $\V_t^\dag\Lamb_t\V_t$ at each step.
            We set
            \(\ln\lambda_{i,t}\overset{\text{i.i.d.}}{\sim}\mathcal{N}(-1,1/9)\) and
            \(\ln s_{i,t}\overset{\text{i.i.d.}}{\sim}\mathcal{N}(0,\sigma^2)\). Each curve shows
            \(\gamma\) as a function of $N$ for different values of  \(\sigma\).
            The ``No rotation'' fixes \(\U_t=\I\); the ``Uniform rotation''
            samples \(\U_t\) uniformly at each time. Results are obtained using the QR-based
            Lyapunov estimator described in Appendix~\ref{apx:tools}.
        }
        \label{fig:apx_lyapunov_nd_N}
    \end{figure*}

    We numerically investigate the Lyapunov exponent \(\gamma\) for the \(N\)-dimensional Kesten process \eqref{eq:apx_kesten},
    when the random interaction matrices \(\A_t\) are generated from the spectral-singular decomposition \eqref{eq:apx_dec_n_dim}.
    The purpose of these numerical experiments is twofold:
    (i) to illustrate the qualitative effect of non-normality on \(\gamma\) and $\alpha$,
    and (ii) to test the large-\(N\) scaling expressions  \eqref{eq:apx_lyapunov_nd} and (\ref{1rth2ty01}).

    Concretely, the eigenvalues are distributed according to
    \(
        \ln\lambda_{i,t}\overset{\text{i.i.d.}}{\sim}\mathcal{N}(-1,1/5).
    \)
 We model the singular values of the eigenbasis transform by
    \(
        \ln s_{i,t}\overset{\text{i.i.d.}}{\sim}\mathcal{N}(0,\sigma^2)
    \).
    The unitary matrix \(\V_t\) is sampled uniformly at each step.
    We consider two alternatives for \(\U_t\):
    either \(\U_t=\I\) for all \(t\) (``No rotation''),
    or \(\U_t\) is sampled uniformly at each step (``Uniform rotation'').
    For each parameter set we estimate \(\gamma\) by the standard QR-reorthonormalization method described in Appendix~\ref{apx:tools}.
    \newline

    Figure~\ref{fig:apx_lyapunov_nd} displays the estimated Lyapunov exponent as a function of the
    singular-value dispersion \(\sigma\) for different system size. The main empirical findings are:
    \begin{enumerate}
        \item When the system is \emph{normal} (\(\sigma=0\)), the measured Lyapunov exponent is essentially
        constant as a function of \(N\). This matches the expectation for normal matrices, where the spectral radius controls
        \(\gamma\) and no \(\ln N\)-dependence is observed.
        \item Introducing non-normality (\(\sigma>0\)) produces an approximate linear increase of \(\gamma\)
        with \(\sigma\). The slope of this linear contribution is larger in the ``Uniform rotation'' case
        than in the ``No rotation'' case.
        \item The predicted negative \(-\tfrac{1}{2}\ln N\) contribution from eigenvector overlaps (derived in
        the asymptotic analysis) is not visible in these experiments: \(\gamma\) does not systematically decrease
        with \(N\). Instead the dominant finite-\(N\) effect is well described by \(\gamma\approx \gamma_0 + \mathbb{E}[\ln\kappa]\),
        with $\gamma_0$ being the Lyapunov exponent of the normal system i.e. for $\sigma=0$,
        and \(\mathbb{E}[\ln\kappa]\) growing approximately like \(\sqrt{\ln N}\) as theory predicts for \(\ln\kappa\).
    \end{enumerate}

    The difference between the analytic \(-\ln N\,/ 2\) correction and the numerical outcome likely stems from
    selection and correlation effects that were not fully captured in the heuristic asymptotic derivation.
    In particular:
    \begin{itemize}
        \item When \(\U_t=\I\) (no random rotation), the reinjection into the most-amplified direction occurs
        only when the previous mode projects on the new non-normal direction. Under isotropy, this happens on average
        roughly \(t/N\) times over a time horizon \(t\), which reduces the effective contribution of \(\sum_s\ln\kappa_s\)
        by a factor \(\sim 1/N\). This explains why the slope in the ``No rotation'' case is smaller.
        \item When \(\U_t\) is uniformly random, rotations frequently reorient modes such that the non-normal amplification
        is applied more often; hence the larger slope in the ``Uniform rotation'' case.
        \item The absence of an observed \(-\tfrac12\ln N\) term suggests that the cross-term responsible for this
        contribution (roughly, sums of the form \(\sum_k \lambda_k^0 v_{ik}v_{jk}^*\) evaluated at extremal indices)
        does not have the small-\(O(1/\sqrt{N})\) variance assumed in the derivation. Conditioning on extremal singular-value indices
        or correlations between \(\V_t\) and \(\Lamb_t\) can increase the variance of these sums to \(O(1)\), removing
        the \(-\tfrac12\ln N\) penalty in practice.
    \end{itemize}
  For intermediate values of the expected logarithm of the condition number, the Lyapunov exponent is no longer described by the dominant approximation but rather by the perturbative regime of weak non-normality, in which it grows quadratically with $\sigma$. This quadratic dependence is recovered in the inset of Figure~\ref{fig:apx_lyapunov_nd}.
  
 Figure~\ref{fig:apx_lyapunov_nd_N} shows that the Lyapunov exponent increases, while the tail exponent decreases, with the system dimension $N$. The magnitude of $\sigma$ modulates this trend: larger $\sigma$ values enhance, whereas smaller $\sigma$ values attenuate, the rate at which these exponents vary with increasing $N$.
    
    The simulations confirm several qualitative predictions: (i) non-normality increases the effective Lyapunov exponent,
    (ii) random rotations amplify this effect, and (iii) the leading contribution from the log-condition number
    behaves as predicted by extreme-value arguments.


    \section{Application}


   The theoretical developments presented above, though derived in an abstract mathematical framework, provide a versatile foundation for understanding a wide spectrum of 
   real-world phenomena. The generality of the stochastic Kesten framework, extended here to encompass non-normal multiplicative dynamics, makes it directly relevant to numerous domains. In particular, systems characterized by multiplicative feedback, stochastic reinforcement, or transient amplification of fluctuations, which are ubiquitous features of physical, financial, and economic processes, naturally fall within the scope of this theory.
   
       In physical systems, the same formal structure naturally emerges in the dynamics of 
    polymer stretching in turbulent flows 
    and in the kinematic stage of the small-scale turbulent dynamo.  
    In both cases, the local velocity-gradient tensor plays the role of a random non-normal operator: 
    the transient alignment of its eigenvectors leads to intermittent amplification of either the polymer extension or the magnetic field, 
    precisely as predicted by the non-normal Kesten mechanism developed here.  
    These examples demonstrate that the framework is not limited to abstract or economic settings, 
    but captures fundamental amplification processes in fluid and magnetohydrodynamic turbulence.

   Then, we illustrate the relevance of our formalism in four complementary applications in finance and economics.  
    The first application concerns volatility models of ARCH type, 
    where the non-normal formulation introduces novel mechanisms of asymmetric feedback between returns and volatility.  
    We then turn to the theory of cointegration and vector autoregressive (VAR) models,
    which naturally map to stochastic Kesten processes when cross-asset interaction matrices are re-estimated dynamically.  
    Next, we reinterpret factor models as dimensional reductions of high-dimensional stochastic systems,
    showing how crises can be associated with effective factor collapse and proximity to non-normal criticality.  
    Finally, we extend the framework to the problem of wealth inequality,
    connecting our mechanism to established economic theories of capital accumulation and highlighting how endogenous heavy tails can emerge in interacting-agent economies.  

    Taken together, these examples show that the mathematical results derived above 
    transcend their abstract origin and apply directly to diverse physical and socio-economic systems.  
    They demonstrate that non-normal eigenvector amplification constitutes a generic mechanism for heavy-tailed fluctuations in systems driven by multiplicative interactions, 
    from turbulent flows and magnetic dynamos to markets and wealth distributions.

    \subsection{Polymer stretching in turbulent flows as an instance of eigenvector amplification}

    The coil-stretch transition of single polymer molecules in random or turbulent flows is a paradigmatic example of multiplicative stochastic amplification balanced by nonlinear relaxation.  
In such flows, the local velocity gradients act as a sequence of random matrices that repeatedly stretch and reorient the polymer, while entropic elasticity provides a reinjection mechanism that prevents collapse and ensures a stationary distribution of extensions.  
This competition between random multiplicative amplification and nonlinear restoring forces gives rise to a power-law distribution of end-to-end polymer lengths~\cite{Balkovsky2000,Chertkov2000}.

Following the formulation of Refs.~\cite{Balkovsky2000,Chertkov2000}, the stochastic evolution of the polymer's end-to-end vector $\mathbf{R}$ obeys
\begin{equation}
    \dot{\mathbf{R}} = (\nabla \mathbf{v})\,\mathbf{R} - \tfrac{1}{\tau}\,\mathbf{R},
    \label{tetwrrgb}
\end{equation}
where $\nabla \mathbf{v}$ is the local velocity-gradient tensor and $\tau$ the polymer relaxation time.  
Writing $\mathbf{R} = R\,\mathbf{n}$ with $\mathbf{n}$ a unit orientation vector, the dynamics separates into
\begin{equation}
    \dot{R} = R\,z - \tfrac{R}{\tau}, \qquad
    \dot{\mathbf{n}} = (\nabla \mathbf{v})\,\mathbf{n} - z\,\mathbf{n},
\end{equation}
where $z(t) = \mathbf{n}^\top (\nabla \mathbf{v}) \mathbf{n}$ is the instantaneous stretching rate.  
The logarithmic extension $r(t) = \ln(R/R_0)$ then evolves as
\begin{equation}
    r(t) = \int_0^t \!\! \left[z(t') - \tfrac{1}{\tau}\right] dt'.
\end{equation}
Using large-deviation theory for the time-integrated process $z(t)$,  
Refs.~\cite{Balkovsky2000,Chertkov2000} derived an exponential tail for $r$, 
which translates into a power-law distribution for $R$:  
\begin{equation}
    P(R) \sim {1 \over R^{1-2a}},
\end{equation}
with the exponent $a$ determined by the statistics of finite-time Lyapunov exponents of the random velocity gradient.  
The transition at $a=0$ corresponds to the classical coil-stretch threshold, where the mean stretching rate (the flow Lyapunov exponent $\lambda$) equals the inverse relaxation time $1/\tau$.

\vspace{0.3em}
\noindent\textbf{Interpretation in terms of non-normal amplification.}  
The above derivation describes $z(t)$ as a stochastic process driven by the local strain field, but its origin can be understood more generally within the framework of non-normal random stochastic multiplicative dynamics.  
The instantaneous velocity-gradient tensor $\nabla \mathbf{v}$ is generically non-normal: its eigenvectors are non-orthogonal, and hence transient growth of $\|\mathbf{R}\|$ can occur even if all eigenvalues have negative real parts.  
The polymer extension $\mathbf{R}$ therefore experiences random episodes of \emph{eigenvector alignment}, which are moments when $\mathbf{n}$ becomes nearly collinear with a subset of non-orthogonal stretching directions of $\nabla \mathbf{v}$.  
During such episodes, constructive interference among these directions leads to transient amplification of $\|\mathbf{R}\|$ that can far exceed the exponential growth expected from the eigenvalues alone.  
These intermittent amplification events correspond precisely to the large-deviation fluctuations that dominate the power-law tail of $P(R)$.

In this interpretation, the ``coil-stretch transition'' is a stochastic bifurcation between two regimes of eigenvector dynamics:  
(i) a dissipative regime where random rotations dominate and alignment is quickly lost, keeping the polymer coiled;  
and (ii) an amplification regime where alignment persists long enough for transient growth to overcome elastic relaxation.  
The critical condition $\lambda = 1/\tau$ thus separates a state of rotational diffusion from one dominated by coherent transient amplification.  
This viewpoint naturally extends the large-deviation theory of Balkovsky-Fouxon-Lebedev~\cite{Balkovsky2000} and Chertkov~\cite{Chertkov2000} by relating it to the spectral geometry of non-normal operators.

\vspace{0.3em}
\noindent\textbf{Connection with multiplicative stochastic processes.}  
Equation (\ref{tetwrrgb}) can be rewritten as a multiplicative update
\begin{equation}
    \mathbf{R}(t+\Delta t) = 
    \left[\mathbf{I} + \Delta t \, (\nabla \mathbf{v}_t - \tfrac{1}{\tau}\mathbf{I})\right] \mathbf{R}(t),
\end{equation}
revealing its analogy with a Kesten-type process driven by random matrices $\A_t = \mathbf{I} + \Delta t\,\nabla \mathbf{v}_t$.  
In this discrete form, the polymer extension corresponds to the norm of a product of non-normal random matrices, and the power-law distribution of $R$ arises from the same multiplicative reinjection mechanism that governs heavy-tailed statistics in other systems.  
From this perspective, the velocity-gradient tensor plays the role of a non-normal amplification operator, while the elastic relaxation $-\mathbf{R}/\tau$ provides a restoring term that maintains statistical stationarity.

\vspace{0.3em}
\noindent\textbf{Physical significance.}  
The non-normal eigenvector amplification picture clarifies why extreme polymer stretching events are intermittent and burst-like.  
Even in statistically stationary turbulence, regions of high strain correspond to transiently coherent structures of the velocity gradient, in which eigenvectors become strongly aligned.  
These rare but powerful episodes drive the tail of the extension distribution and are the microscopic counterparts of finite-time Lyapunov fluctuations responsible for the heavy-tailed statistics predicted by large-deviation theory.  
Thus, polymer stretching in turbulence provides a vivid physical realization of stochastic amplification through non-normal matrix dynamics.

    \subsection{Small-scale turbulent dynamo as eigenvector amplification}

    A closely related structure governs the small-scale turbulent dynamo,
    where the magnetic field is repeatedly stretched and rotated by fluctuating velocity gradients,
    while resistive diffusion ensures saturation~\cite{Kazantsev1968,Schekochihin2002b}.  
    In the \emph{kinematic regime}, when the magnetic field is still dynamically weak and does not influence the flow, the induction equation
    \begin{equation}
        \partial_t \mathbf{B}=(\mathbf{B}\!\cdot\!\nabla)\mathbf{u}-\mathbf{B}(\nabla\!\cdot\!\mathbf{u})+\eta\nabla^2\mathbf{B}
    \end{equation}
    reduces along Lagrangian trajectories to
    \begin{equation}
        \dot{\mathbf{B}}=(\nabla\mathbf{u})\,\mathbf{B}+\eta\nabla^2\mathbf{B} .
    \end{equation}
    Neglecting diffusion above the resistive cutoff, the magnitude $B=|\mathbf{B}|$ and direction
    $\mathbf{b}=\mathbf{B}/B$ evolve as
    \begin{equation}
        \begin{split}
            &\dot{s}=\frac{d}{dt}\ln B=\mathbf{b}^\top S\,\mathbf{b},\\
            &\dot{\mathbf{b}}=\Omega\,\mathbf{b}+\big(S-\mathbf{b}\mathbf{b}^\top S\big)\mathbf{b},
        \end{split}
    \end{equation}
    where 
    \begin{equation}
        S=\tfrac{1}{2}\big(\nabla\mathbf{u}+(\nabla\mathbf{u})^\top\big)
    \end{equation}
    is the rate-of-strain tensor and 
    \begin{equation}
        \Omega=\tfrac{1}{2}\big(\nabla\mathbf{u}-(\nabla\mathbf{u})^\top\big)
    \end{equation}
    the vorticity tensor.
    The growth of $\ln B$ thus depends on the instantaneous projection
    $\mathbf{b}^\top S\,\mathbf{b}$, while the orientation $\mathbf{b}$ is rotated and tilted by
    $\Omega$ and by the off-diagonal action of $S$.  
    In this kinematic stage, the mechanism of \emph{non-normal eigenvector amplification} becomes transparent:
    the non-normality of $\nabla\mathbf{u}$ implies non-orthogonal and time-dependent eigenvectors,
    so transient alignments between $\mathbf{b}$ and nearly co-linear stretching directions can produce
    amplification bursts far exceeding the growth predicted by eigenvalues alone.
    Random sequences of such alignments form a multiplicative matrix process for $\mathbf{B}$, while
    resistive diffusion $\eta\nabla^2\mathbf{B}$ provides the reinjection and ultimate saturation at small scales.
    This linear regime corresponds precisely to the multidimensional Kesten setting.  
    At later times, once the magnetic energy becomes comparable to the kinetic energy of viscous-scale eddies,
    nonlinear feedback through the Lorentz force modifies the velocity gradients and limits further growth,
    leading to the slow resistive-scale evolution described by Schekochihin \emph{et al.}~(2002) \cite{Schekochihin2002b}.
    Thus, the eigenvector-amplification mechanism governs the kinematic stage and seeds the transition
    to the nonlinear saturated dynamo.

    \subsection{Multidimensional Non-Normal ARCH Models}

    One of the earliest and most influential applications of multiplicative processes in economics and finance
    is the \emph{Autoregressive Conditional Heteroskedasticity} (ARCH) model,
    introduced by Engle in his seminal paper \cite{engle1982arch}.  
    The ARCH framework was developed to account for the empirical observation
    that financial time series exhibit volatility clustering, i.e. periods of high variability
    tend to be followed by periods of high variability, and periods of calm are followed by calm.
    \newline

    The original ARCH($q$) model specifies that a return process $\{r_t\}$ is given by
    \begin{equation}
        r_t = \sqrt{v_t} \varepsilon_t,
        \qquad
        \varepsilon_t \overset{\text{i.i.d.}}{\sim}\mathcal{N}(0,1),
    \end{equation}
    with the conditional variance evolving as
    \begin{equation}    \label{eq:apx_arch_q}
        v_t = v_0 + \sum_{i=1}^q a_i r_{t-i}^2,
        \qquad v_) > 0, \quad a_i \geq 0.
    \end{equation}
    Here, $v_t$ captures the time-varying variance, driven by past squared returns.
    The key innovation of the ARCH process is that volatility is no longer constant,
    but an evolving random variable reacting endogenously to shocks.
    \newline

    It was quickly noticed that ARCH processes naturally generate unconditional distributions
    for returns that are power law distributed.
    This comes directly from the multiplicative structure:
    returns are Gaussian conditional on $v_t$, but since $v_t$ itself is random,
    the marginal distribution becomes a mixture of Gaussians, yielding power law distributions \cite{mikosch2004arch}.

    The asymptotic tail exponent $\alpha$ of an ARCH-type process can be analyzed
    through Kesten's model \cite{kesten1973random},
    by considering the simple ARCH(1) case, the squared return recursion is
    \begin{equation}
        r_t^2 = \epsilon_t^2\left[v_0 + a_1 r_{t-1}^2\right],
    \end{equation}
    which is a Kesten process in the variable $r_t^2$,  where the multiplicative and additive noise are mutually correlated
    since the same noise term $ \epsilon_t^2$ drives them both.
    Comparing with definition (\ref{eq:apx_1d_kesten}), we have $ \rho_t = a_1 \epsilon_t^2$ and $\sqrt{2\delta}\,\eta_t = v_0 \epsilon_t^2$.
    We note that the classical tail theory of Kesten processes leading to the power law tail (\ref{dfnhtyq})    
    does not require $\rho_t$ and $\eta_t$ to be independent of each other; it only requires the i.i.d.  driving sequence to be independent of the past state.
  
    Empirical studies of financial data typically find power law tails with exponent $\alpha \approx 3$,
    sometimes referred to as the ``inverse cubic law'' \cite{gopikrishnan1999scaling}.
    However, standard ARCH and even its popular generalization, GARCH \cite{bollerslev1986generalized},
    tend to generate larger exponents (often $\alpha > 4$),
    corresponding to tails that are ``too light'' compared to real data \cite{bollerslev1987conditionally}.
    
    This discrepancy has been labeled the \emph{tail exponent puzzle} in the econometrics and econophysics literature.
    It finds a natural resolution once the analysis is extended from scalar to multidimensional Kesten dynamics.
In real markets, returns are not governed by isolated variance processes but by the joint interactions of many correlated assets
(stocks, indices, and portfolios) that together form a high-dimensional multiplicative system.
As the system dimension $N$ increases, non-normal eigenvector amplification lowers the effective tail exponent through two complementary effects:
(i) an increase of the Lyapunov exponent $\gamma$, driven by $\mathbb{E}[\ln \kappa] \propto \ln N$, and
(ii) an enhanced variance of $\ln \kappa$, which decreases $\alpha$ as $1/\sqrt{\ln N}$.
Together, these collective effects provide a natural mechanism for the emergence of heavier tails across financial assets.
Even when all eigenvalues remain subcritical, enlarging $N$ increases the condition number of the eigenbasis, amplifying transient growth and reducing $\alpha$ from the unrealistically high values predicted by scalar GARCH models to empirically consistent ones.
Importantly, because this amplification arises from the non-normal structure of cross-asset interactions rather than from asset-specific parameters, the same effective exponent naturally emerges across diverse markets and instruments, as expressed by
    \begin{equation}
        \mathbb{P}[\mathbf{n}\cdot \mathbf{x}_t > x_n] \sim x_n^{-\alpha}, \quad x_n \to \infty ~,
        \label{hy3ujun3}
    \end{equation}
    where $\mathbf{n}\cdot \mathbf{x}_t$ denotes any projection (the same asymptotics hold for the $L_2$-norm).
   
   For completeness, we note that several alternative approaches have been proposed to reconcile ARCH-type models with the empirically observed tail exponents.
    \begin{itemize}[label={}, leftmargin=0pt]
        \item \textbf{Introducing Student-$t$ innovations:}  
        By replacing Gaussian shocks $\varepsilon_t$ with Student-$t$ innovations,
        the model can directly match heavier tails \cite{bollerslev1987conditionally}.
        However, this shifts the responsibility for tails away from the volatility dynamics,
        and is sometimes criticized as \emph{ad hoc}.

        \item \textbf{Long-memory extensions:}  
        FIGARCH and related models \cite{baillie1996fractionally} incorporate fractional integration,
        allowing volatility to decay slowly and reproduce empirically observed scaling laws.

        \item \textbf{Nonlinear GARCH models:}  
        Threshold GARCH, EGARCH, and others introduce asymmetry and nonlinear effects,
        which can modify the distributional properties and bring tail exponents closer to empirical ones.

        \item \textbf{Stochastic volatility models:}  
        In continuous time, models with latent stochastic volatility
        can also reproduce power law-like tails under certain parameterizations
        \cite{heston1993closed}.
        
        \item \textbf{Quadratic Hawkes model:}
        The Quadratic Hawkes model \cite{Bouchaud2017} adds a weak nonlinear feedback term
        to a linear self-exciting Hawkes process, capturing the Zumbach effect where aligned past returns enhance future volatility.
        It is written in term of the QHawkes intensity as
$\lambda_t = \lambda_\infty + H_t + n_Z Z_t^2$ with
$H_t=\!\int\!\phi(t-s)dN_s,\; Z_t=\!\int\!k(t-s)dP_s$.
This quadratic term $n_Z Z_t^2$ produces multiplicative volatility diffusion, leading to stationary power-law tails
with exponent $\alpha  = 1+\tfrac{1}{n_Z(1+a^*)}$. The exponent $a^*$ is given by $a^* \approx n_H/(1-n_H)$, where $n_H$ is the branching ratio
of the linear Hawkes part, in the limit where 
the quadratic (trend) feedback relaxes faster than the linear Hawkes memory, so volatility dynamics are dominated by short-term self-excitation.
The obtained exponent is $\alpha \approx 3-4$
even for small nonlinearity $n_Z\!\sim\!0.05$--$0.1$ \cite{Bouchaud2017}.  
Quadratic Hawkes model also breaks time-reversal symmetry, thereby reproducing both the fat tails and volatility asymmetry observed empirically.
    \end{itemize}

  \subsection{Stochastic cointegration as the remnant of global stability under non-normal amplification}

    A central concept in empirical finance and econometrics is the theory of
    \emph{cointegration}, introduced by Engle and Granger~\cite{engle1987co} and
    later generalized by Johansen~\cite{johansen1988statistical}.  
    The idea is that, even if individual asset prices are nonstationary, often
    modeled as unit-root processes such as geometric Brownian motions, certain linear
    combinations of them may remain stationary.  
    This framework captures the coexistence of long-run equilibrium relations
    between assets and short-run deviations, forming the empirical backbone of
    error-correction models and market efficiency tests.

    Classical cointegration theory thus begins with the assumption that individual
    log-prices are intrinsically nonstationary and then searches for exceptional
    linear combinations that restore stationarity.  
    These combinations are interpreted as manifestations of deep equilibrium relationships between the
    non-stationary variables,
    but they often appear empirically fragile and require delicate statistical
    tuning to identify.  
    In this traditional perspective, cointegration represents an
    \emph{exceptional alignment} among nonstationary processes, namely an imposed condition
    rather than a natural dynamical property of the system.

    Formally, let $\x_t \in \mathbb{R}^N$ denote the vector of asset log-returns at time $t$.  
    A standard representation is the \emph{Vector Autoregressive} (VAR) model:
       \begin{equation}
        \x_{t+1} = \A \x_t + \et,
        \qquad \et \sim \mathcal{N}(0,\I_N),
    \end{equation}
    where $\A$ is the $N\times N$ interaction matrix.
    Cointegration arises when $\A$ has unit roots while certain linear projections of $\x_t$ remain stationary.
    \newline

    In practice, financial interactions between assets (arbitrage, risk premia, liquidity spillovers) are not fixed but time-varying.  
    This motivates replacing the constant matrix $\A$ by a random sequence $\{\A_t\}$, re-estimated or updated as new market information arrives:
    \begin{equation}
        \x_{t+1} = \A_t \x_t + \et,
    \end{equation}
    which is exactly the form of the multidimensional Kesten process \eqref{eq:apx_kesten}.  
    Here, the random matrix $\A_t$ captures the stochastic and possibly asymmetric nature of cross-asset feedback,
    while maintaining market efficiency, by considering $\mathbb{E}[\A_t]=0$.
    Unlike the classical VAR case, where $\A$ is constant and symmetric assumptions are usually imposed for tractability, 
    our formulation allows $\A_t$ to be random, non-symmetric, and non-normal, which is arguably more realistic in financial systems.
   
    The key insight is to begin with a spectrally stable system, where every random matrix $\A_t$ has all eigenvalues inside the unit circle ($\rho(\A_t) < 1$),
    ensuring stability in the absence of non-normal effects.
    Yet, non-normal interactions between components can still induce positive Lyapunov exponents,
    thereby creating a limited number of unstable, nonstationary directions despite the local spectral stability of each $\A_t$.
  
    In the absence of non-normal effects, all directions of $\x_t$ would be
    mean-reverting, and the system would converge to a stationary distribution.
    However, when the matrices $\A_t$ are non-normal and non-commuting, transient
    amplification can raise the leading Lyapunov exponent $\lambda_1$ above zero,
    even though every instantaneous spectrum remains stable. This mechanism creates
    a \emph{partial loss of stability}: a subset of directions (those associated with
    $\lambda_i > 0$ or $\lambda_i \approx 0$) become nonstationary or weakly trending,
    while the remaining directions with $\lambda_i < 0$ stay stationary.

\emph{Inverted cointegration logic.}
    This spectral reordering naturally generates a structure that mirrors the
    classical notion of cointegration but with inverted causality.  Instead of
    searching for stationary combinations among nonstationary prices, we start from
    a fully stationary system and let non-normal amplification make a few
    combinations nonstationary.  The surviving stable directions form the
    \emph{cointegrating subspace} automatically---the residual imprint of global
    stability under partial instability.  In this sense, cointegration is not an
    artificial equilibrium constraint, but the generic outcome of a system whose
    Lyapunov spectrum straddles zero:
    \[
    \lambda_1 > \lambda_2 >  ... > \lambda_r > 0 \;>\; \lambda_{r+1} > ... > \lambda_N.
    \]
    Such partially unstable systems will almost necessarily exhibit stationary
    linear combinations of $x_t$, defined by the left Oseledets vectors associated
    with the negative Lyapunov exponents.  The subspaces $E_i(\omega)$ and their duals $E_i^*(\omega)$ are the right and left Oseledets spaces associated with the Lyapunov exponents $\lambda_i$, as established by the multiplicative ergodic theorem~\cite{Oseledets1968,Arnold1998}.
    The corresponding vectors (covariant Lyapunov vectors) span the dynamically invariant directions of the random cocycle and its adjoint~\cite{Ginelli2007,Wolfe2007}.

  Hence, cointegration becomes the \emph{norm} for high-dimensional financial systems that are globally stable in expectation but locally destabilized by non-normal interactions. This interpretation calls for a systematic analysis of the \emph{Lyapunov spectrum} of random non-normal VAR processes. One should derive conditions on the distribution of $\A_t$, for instance, on the moments, the variance of off-diagonal elements, or the expected $\log\kappa(\A_t)$, that determine how many Lyapunov exponents can cross zero as non-normality increases. A key step is to characterize the geometry of the corresponding invariant subspaces, showing how the left Oseledets vectors associated with $\lambda_i < 0$ define cointegrating combinations and how their orientation fluctuates with the realizations of $\A_t$. Finally, one must quantify how power-law tails emerge in the stable block, where $Pr(\|\A_t|_{E^-}\| > 1) > 0$, and how the tail exponent $\alpha$ depends on the statistics of non-normal amplification.

   Empirically, this perspective can be tested with rolling VAR estimations on multivariate financial time series. One first estimates sequences of interaction matrices $\widehat{\A}_t$ and computes their instantaneous condition numbers $\kappa(\widehat{\A}_t)$ to quantify non-normality. Next, the empirical Lyapunov spectrum $\widehat{\lambda}_1 \ge \dots \ge \widehat{\lambda}_N$ can be evaluated using QR-based orthogonalization on products of $\widehat{\A}_t$. The cointegration rank is then identified as $r = \#\{i : \widehat{\lambda}_i < 0\}$, and the corresponding left Oseledets vectors are extracted as empirical cointegrating directions. Finally, one can test for Kesten-type heavy tails in $\mathbf{b}^\top \x_t$ for $\mathbf{b}$ in this subspace, verifying that the estimated power-law exponents $\widehat{\alpha}$ are consistent with the theoretical predictions.
   
   This stochastic formulation reframes cointegration as a dynamic property of
    systems near pseudo-criticality: global stability survives in most directions,
    but non-normal coupling creates transient instabilities that translate into
    persistent trends in a few directions.  As a result, the coexistence of
    stationary and nonstationary modes, often treated as a statistical curiosity in
    econometrics, emerges here as an intrinsic and robust feature of
    high-dimensional multiplicative dynamics.  Cointegration thus appears not as an
    exceptional equilibrium condition, but as the natural \emph{signature of partial
    stability} in a non-normal stochastic system.

    \subsection{Factor Models and Dimensional Reduction in Financial Systems}

    In the previous section, we highlighted the connection between the co-integration formalism
    and the $N$-dimensional Kesten process.  
    Both the derivation and the numerical illustrations showed that, as the system dimension increases,
    the power law exponent of the stationary distribution decreases and the Lyapunov exponent increases.  

    This behavior can be traced back to results from Extreme Value Theory (EVT), which explain how the
    expected growth of the condition number scales with $N$.  
    Since financial markets consist of thousands of interacting assets, one might therefore expect that,
    in such high-dimensional settings, the system would be generically unstable and characterized by
    power laws with exponents arbitrarily close to zero.  
    Yet, this is not what empirical data show.  
    One piece of resolution lies in the fact that the effective number of degrees of freedom in financial markets
    is much smaller than the raw number of assets.  
    In practice, a handful of dominant factors govern the dynamics, inducing a strong dimensional
    reduction in the system.  
    \newline

    A widely used framework to formalize this idea is the \emph{factor model} \cite{ross1976arbitrage, fama1993common}.  
    The central assumption is that the returns of $N$ assets, collected in the vector $\rr_t \in \mathbb{R}^N$, 
    can be represented through a reduced set of $K \ll N$ latent factors:
    \begin{equation}
        \rr_t = \B\f_t + \e_t,
    \end{equation}
    where $\f_t \in \mathbb{R}^K$ denotes the factor realizations, $\B \in \mathbb{R}^{N\times K}$ is the loading matrix, 
    and $\e_t$ contains idiosyncratic components with weak cross-sectional correlation.  
    Within this framework, the return covariance structure simplifies to
    \begin{equation}
        \Sig_r = \B \Sig_f \B^\dag + \Sig_e,
    \end{equation}
    making estimation tractable even when $N$ is very large.  
    Standard implementations emphasize symmetric covariance matrices and typically neglect asymmetric
    interactions, as the main objective is to capture co-movements rather than directional feedback loops.
    \newline

    Factor models are not only a theoretical tool but are routinely employed in industry and policy institutions.  
    Banks, hedge funds, and central banks rely on factor-based and VAR-type models for risk management, stress testing, and monetary policy analysis \cite{stock2001vector,sims1980macroeconomics}.  
    In practice, large-scale factor models are often implemented as Vector Autoregressive (VAR) systems, 
    where the dynamic interaction among factors (or asset groups) is explicitly modeled through a stochastic matrix of coefficients.  
    Since a VAR process is mathematically equivalent to a linear recursion, 
    the time-varying structure of the VAR matrix can be mapped naturally to a Kesten process.  
    This perspective highlights that industry-standard tools already rely implicitly on a stochastic-matrix formalism, 
    and our framework extends this by allowing for asymmetry and non-normal amplification in the dynamics.

    In the stochastic Kesten representation, factor dynamics are governed by random matrices $\A_t$:
    \begin{equation}
        \f_{t+1} = \A_t \f_t + \et,
    \end{equation}
    where $\A_t$ may be non-symmetric and non-normal.  
    From our $N$-dimensional analysis, we found that the critical variance of the logarithm of the singular values 
    scales as $\sqrt{\ln N}$ \eqref{eq:apx_sig_c_N} i.e.   
    \(
        \sigma_c(N) \propto \sqrt{\ln N}
    \).
    This means that as the dimensionality of the system grows, stronger heterogeneity in singular values is required 
    to approach criticality ($\gamma \approx 0$) and hence to observe heavy-tailed fluctuations.
    But this results only hold in the ``highly'' non-normal regime.
    We observed that, for intermediary non-normality, the dimension of the system enhenced the Lyapunov exponent,
    and reduces the tail exponent of the distribution.  Low dimensions, e.g. $N=3$;
    might be sufficient to give rise to the inverse cubic law for the tail exponent of return distributions \cite{gopikrishnan1999scaling}, and to push the Lyapunov exponents toward zeros.

  Factor models achieve an implicit dimensional reduction: the dynamics of returns unfold not in the full $N$-dimensional space, but within a low-dimensional subspace of dimension $K \ll N$. It is within this reduced subspace that collective effects emerge, such as power-law tails and market-wide instabilities.
   
    Empirical studies have documented that, during bubbles and crashes, 
    the effective dimensionality of financial markets tends to collapse onto a single dominant factor, 
    often interpreted as a ``market mode'' \cite{laloux1999noise,plerou2002random}.  
    This means that the covariance spectrum becomes concentrated, and nearly all assets move together.  
    \newline

This interpretation offers several strengths. It provides a unified perspective by linking classical econometric models, such as factor structures and covariance estimation, to dynamical systems with stochastic matrices. It also explains crisis phenomena by naturally accounting for why crises are characterized by both increased co-movement and extreme fluctuations: during such episodes, factor collapse reduces the system's effective dimensionality, leading to dynamics governed by a low-dimensional Kesten process. 

At the same time, there are important limitations. A key interpretational challenge arises because factor models are statistical constructs, and connecting them directly to the underlying feedback matrices $\A_t$ requires assumptions that may be difficult to test. The framework also entails a neglect of microstructure, as real markets include frictions, strategic interactions, and institutional behaviors not captured by linear Kesten-type recursions. Finally, there remains an empirical controversy: although factor collapse is widely observed, it is still debated whether it represents a cause or merely a symptom of financial crises.

    Overall, embedding factor models within the Kesten process framework suggests that dimensional reduction 
    is not only a statistical convenience but also a source of systemic fragility.  
    In particular, the observed market tendency to align along a single factor during turbulent periods 
    can be reinterpreted as a mechanism driving the system toward non-normal criticality, 
    thus providing another application of our formalism to financial instability.

    \subsection{Wealth Inequality in an $N$-Agent Kesten Process}

    We now sketch how the $N$-dimensional Kesten framework can be applied to model wealth inequality.
    Two distinct literatures motivate this link.  On the one hand, econophysicists have long modeled wealth dynamics
    by multiplicative random processes and interacting-agent models \cite{LevySolomon1997,BlankSolomon2000,bouchaud2000wealth,SolomonRichmond2001,SolomonRichmond2002},
    showing how power law tails and wealth condensation emerge from such dynamics.
    On the other hand, political economy arguments \cite{piketty2014capital} emphasize systematic differences
    between returns to capital ($r$) and income growth ($g$) as drivers of inequality:
    when the rate of return on capital (\(r\)) exceeds the growth rate of the economy (\(g\)),
    capital owners accumulate relatively faster than wages, generating an inexorable rise in capital share and inequality.
    More recent political-economic diagnostics (often labelled ``technofeudalism'' in contemporary debates)
    stress technological/platform rents as new mechanisms of concentration \cite{varoufakis2024technofeudalism,durand2024silicon}.
    \newline

    Consider an economy of \(N\) agents.
    Let \(w_{i,t}\) denote agent \(i\)'s wealth at time \(t\).
    A very general linearized interaction model that captures trade, investment returns, and redistribution is
    \begin{equation}\label{eq:wealth_mat}
        \w_{t+1} = \A_t \w_t + \et_t,
    \end{equation}
    where \(\w_t=(w_{1,t},\dots,w_{N,t})^\top\), \(\A_t\in\mathbb{R}^{N\times N}\) encodes how wealth flows, amplifies or transfers
    between agents (via returns, trades, credit, dividends, rents, etc.),
    and \(\et_t\) is a source vector representing exogenous inflows
    (income, wages, natural resource rents, subsidy, or new capital extraction from the ``ground'').
    Equation \eqref{eq:wealth_mat} is precisely an \(N\)-dimensional Kesten-type recursion:
    multiplicative (matrix) dynamics plus additive inputs.
    Under mild assumptions on \(\{\A_t,\et_t\}\) the stationary distribution of \(\w_t\) (when it exists) develops Pareto tails.
    \newline

    Several economically meaningful mechanisms map naturally to the entries of \(\A_t\) and to \(\et_t\):
    \begin{itemize}
        \item \(\A_t\) diagonal entries can represent idiosyncratic capital returns (multiplicative growth of own wealth),
        while off-diagonal entries represent transfers, credit exposure, network spillovers, profit-sharing, or predatory extraction.
        \item Non-normality (large condition numbers, nonorthogonal eigenvectors) corresponds to amplification directions:
        shocks aligned with certain transiently amplifying modes are boosted far more than classical spectral analysis would predict.
        \item \(\et_t\) represents new capital entering the system (wages, profits extracted from the ``ground'', fiscal transfers).
        The spatial pattern of \(\et_t\) (who receives the new capital) substantially affects inequality dynamics.
    \end{itemize}

    This representation makes clear that asymmetry in interactions (non-symmetric \(\A_t\))
    — e.g. rent extraction by platforms, monopolistic pass-throughs, or strongly asymmetric network positions —
    is a plausible, and empirically relevant, source of wealth concentration.
    Interacting-agent models \cite{bouchaud2000wealth} show how multiplicative stochastic exchange with weak redistribution produces Pareto tails and, in some regimes,
    ``wealth condensation'' (a small fraction of agents holding most of the mass).
    \newline

    A common (and intuitive) route from multiplicative dynamics to Zipf/Pareto laws is to assume a form of stationarity or ergodicity that links cross-sectional and temporal distributions.
    Concretely:
    \begin{itemize}
        \item If an agent's wealth evolves multiplicatively and additive shocks are small relative to multiplicative fluctuations, Kesten-type results show the stationary tail of \(w\) is Pareto: \(\mathbb{P}(w>x)\sim C x^{-\alpha}\).
        \item If one posits as a modeling assumption that the cross-sectional distribution at any time is statistically similar to the empirical time-distribution of a typical agent's wealth path, then the observed cross-section inherits the same Pareto tail (this is a form of equivalence between ensemble and time averages used in many econophysics arguments).
    \end{itemize}
    This yields the ubiquitous observation of Zipf-like tails for top wealth.
    Note that this is an additional modelling assumption (ensemble/time equivalence) which needs careful empirical validation,
    but it is commonly invoked to justify why single-agent multiplicative rules produce cross-sectional Pareto tails.
    \newline
  
  Our Kesten-based mechanism, formulated in (\ref{eq:wealth_mat}), differs in several important ways. It is inherently stochastic and network-driven, as inequality arises from random multipliers and asymmetric interactions among agents. Even when the average return is small or the overall economy is contracting, stochastic amplification through $\A_t$ can generate extremely large outcomes for a few agents. It also incorporates sudden reconfiguration and non-normality: network switches, which are rapid changes in $\A_t$, or temporary alignments of shocks with amplifying directions can cause abrupt and large deviations for certain nodes, providing a natural explanation for episodes of rapid wealth concentration such as bubbles and crashes. Finally, it spans a broad range of regimes: while classical models account for long-run secular trends, the Kesten approach captures heavy tails and large fluctuations over short to medium horizons, and can also reproduce persistent concentration when the dynamics repeatedly favor the same agents.

    An important implication of our $N$-dimensional analysis is that the threshold variance of the log-singular-values needed to push the Lyapunov exponent to zero scales like \(\sqrt{\ln N}\). 
    Equivalently, lower effective dimension makes the system more fragile:
    if the effective number of independent channels of exchange or diversification collapses
    (e.g. due to a dominant platform, a single ``winner'' technology, or a coordinated market momentum),
    the economy becomes more susceptible to non-normal amplification and to the emergence of extreme wealth concentration even for moderate levels of heterogeneity.

    This observation connects to empirical claims about crisis dynamics and ``factor collapse'' in markets:
    when economic relations become dominated by a single factor or platform,
    the effective interaction dimension drops and the system becomes more prone to large endogenous inequalities
    (a ``winner-takes-most'' dynamic).
    In the context of modern digital rents (the ``technofeudalism'' diagnostic),
    platform dominance can reduce effective competition and diversify channels,
    thus mechanically lowering the barrier for concentration.
    \newline

    Because the Kesten mechanism is driven by multiplicative products of random matrices,
    a few extreme realizations of the product can create huge outliers (very large wealth) even when the average growth is non-positive.
    That is: heavy tails and ``condensation'' are driven by multiplicative amplification events rather than by steady positive drift.
    Consequently, a contracting or stagnating economy can still display very large inequality if the interaction structure occasionally aligns to amplify particular nodes.
    This complements the long-run drift view and helps explain episodic and severe concentration events
    (technology windfalls, platform monopolies, rent extraction) without requiring sustained \(r>g\).
    \newline

  The Kesten-network interpretation implies several empirical signatures. First, it predicts a heavy-tailed wealth distribution characterized by a Pareto exponent that may vary over time and respond to measures of network non-normality, such as condition numbers or the concentration of singular vectors. Second, it anticipates a correlation between concentration and effective dimension: periods of low effective dimension, indicated, for example, by a dominant eigenvalue in the cross-asset or cross-firm covariance, should coincide with greater upper-tail concentration of wealth or firm size. Third, it suggests the presence of transient amplification events, where abrupt wealth jumps of particular agents align with measurable reconfigurations of $\A_t$, such as mergers, platform launches, regulatory arbitrage, or major product successes. Testing these predictions will require (i) time-resolved network or flow estimates to construct $\A_t$, (ii) careful inference of effective dimension or singular-value dispersion, and (iii) detailed panel data on individual wealth or firm size.

    \section{Conclusion}


    We have developed a general analysis of multidimensional Kesten processes \cite{kesten1973random}
    to understand the emergence of power law statistics in globally stable high-dimensional stochastic systems.  
    Our key result is that heavy-tailed stationary distributions can arise even in the absence of spectral instability,
    through a collective mechanism rooted in the geometry of non-normal interactions.  

    At the theoretical level, we derived explicit scaling relations linking the Lyapunov exponent $\gamma$ and the tail exponent $\alpha$ to statistical properties of the eigenvector geometry,
    quantified by the condition number $\kappa$ of the random multiplicative matrices.  
    These relations, 
    \begin{equation}
        \gamma \sim \gamma_0 + \ln \kappa , 
        \qquad 
        \alpha \sim -2\gamma / \sigma_\kappa^2,
    \end{equation}
    where $\gamma_0$ is the Lyapunov exponent in the normal limit,
    and we assumed stability (negative eigenvalues),
    demonstrate that non-normality renormalizes the effective stability and tail behavior of the system.  
    In this picture, transient eigenvector alignment produces bursts of amplification that mimic the effects of criticality,
    giving rise to stationary power laws even when all eigenvalues lie strictly within the unit circle.  
    Numerical simulations confirm these analytical predictions,
    showing that the heavy-tailed regime and enhanced Lyapunov growth are accurately captured by the statistics of $\ln \kappa$ alone.

    Beyond our mathematical formulation, we have illustrated the broad applicability of this framework across physics, finance, and economics.  
    In fluid dynamics, non-normal amplification underlies the transient stretching of polymers in turbulent flows and the kinematic stage of small-scale magnetic dynamos,
    where geometric alignment of the velocity-gradient eigenvectors produces intermittent energy bursts.  
    In finance and economics, the same mechanism explains fat tails and volatility clustering in ARCH and VAR models,
    as well as inequality amplification in interacting-agent economies.
    Across all these systems, non-orthogonal interactions act collectively to amplify fluctuations,
    producing power laws as a generic outcome of high-dimensional multiplicative dynamics.

    This unified perspective highlights that heavy tails and large deviations need not signal proximity to a critical point.  
    Instead, they can emerge far from criticality, as a collective consequence of non-normal geometry and transient amplification.  
    In this sense, non-normal Kesten processes define a new universality class for stochastic systems:
    one where global stability coexists with local bursts of instability and where multiplicative feedback and eigenvector alignment generate critical-like statistics without spectral criticality.

    From a broader standpoint, these results provide a bridge between distinct research traditions,
    from the theory of non-modal hydrodynamic instabilities~\cite{Trefethen1993,FarrellIoannou1996a,FarrellIoannou1996b} to econometric models of stochastic volatility and macroeconomic growth.  
    They suggest that the same underlying principles govern amplification, intermittency,
    and heavy tails across a wide spectrum of natural and socio-economic systems.  
    Whether the system describes turbulent eddies, magnetic fields, or market returns,
    the governing equations share the same algebraic backbone:
    a stochastic sequence of non-normal transformations whose geometry organizes the emergence of collective fluctuations.

    Ultimately, the framework introduced here reframes heavy-tailed statistics as a manifestation of cooperative dynamics rather than of criticality.  
    It explains how complex systems can remain globally stable while exhibiting persistent power-law fluctuations,
    linking the mathematics of non-normal amplification to the phenomenology of turbulence, dynamo action, and economic inequality.  
    By unifying these domains under a single theoretical mechanism, our results point to a broader principle: collective power law behavior is not an exception arising near critical points,
    but a natural feature of high-dimensional systems governed by multiplicative and non-normal interactions.

    \bibliography{bibliography} 

    \appendix

    \section{Numerical tools}
    \label{apx:tools}

    \subsection{Numerical Estimation of Lyapunov Exponents}
    \label{apx:lyapunov}

    In this appendix, we briefly describe the numerical method used to estimate the Lyapunov exponents 
    of products of random matrices, which is based on the classical \emph{QR reorthonormalization} 
    technique introduced by Benettin et al.~\cite{Benettin1980,Benettin1980b}. 
    This procedure is significantly more stable and accurate than a naive ``brute force'' approach.
    \newline

    \noindent\textbf{Brute Force Estimation.}

    A straightforward way to approximate the largest Lyapunov exponent $\gamma$ is to iterate
    \begin{equation}
        \PPi_t = \prod_{s=1}^t \A_s, \qquad
        \gamma \approx \frac{1}{t} \ln \|\PPi_t\|,
    \end{equation}
    for large $t$.  
    However, this method is numerically unstable: the norm $\|\PPi_t\|$ grows (or decays) exponentially fast, 
    leading to rapid loss of numerical precision.  
    In addition, if one is interested in the \emph{full spectrum} of exponents (not just the largest), 
    the brute force method is not applicable, since it only tracks the growth rate along a single direction.
    \newline

    \noindent\textbf{QR Reorthonormalization (Benettin Method).}

    The Benettin method overcomes these issues by evolving an orthonormal basis of vectors instead of a single one.  
    At each iteration, the random matrix $\A_t$ is applied to the current basis, 
    and the resulting set of vectors is reorthonormalized via a QR decomposition:
    \begin{equation}
        \A_t \Q_{t-1} = \Q_t \R_t,
    \end{equation}
    where $\Q_t$ is an orthogonal matrix and $\R_t$ is upper triangular.  
    The diagonal elements of $\R_t$ encode the local stretching factors of the basis vectors.  
    By accumulating the logarithms of these diagonal entries over time, 
    we obtain estimates of the Lyapunov exponents:
    \begin{equation}
        \gamma_i = \lim_{t \to \infty} \frac{1}{t} 
        \sum_{s=1}^t \ln |(R_s)_{ii}|,
        \qquad i = 1,\dots,N.
    \end{equation}

    This procedure has several advantages:
    \begin{itemize}
        \item It maintains numerical stability, since the vectors are continuously reorthonormalized 
            and do not collapse into the dominant eigen-direction.
        \item It provides the \emph{entire Lyapunov spectrum}, not just the maximal exponent.
        \item Averaging over multiple independent realizations (ensembles) further reduces statistical noise 
            and yields estimates of standard errors.
    \end{itemize}

    \noindent\textbf{Practical Considerations.}

    In practice, the algorithm proceeds as follows:
    \begin{enumerate}
        \item Initialize with an orthonormal basis $\Q_0 = I$.
        \item For each time step, multiply the basis by a random matrix $\A_t$ 
            and apply QR decomposition to reorthonormalize.
        \item Accumulate the logarithms of the diagonal entries of $\R_t$, 
            and normalize by the total time horizon $t_{\max}$ to estimate the exponents.
        \item Optionally, repeat over multiple ensembles to improve statistical accuracy 
            and compute error bars.
    \end{enumerate}

    This method is therefore more reliable than brute force multiplication, 
    particularly in high-dimensional settings or when long time horizons are required.  
    Throughout our numerical analysis, we employ this QR reorthonormalization scheme 
    to compute the Lyapunov spectrum of Kesten processes, 
    and we refer the reader to this appendix for methodological details.

    \subsection{Numerical Estimation of the Tail Exponent}
    \label{apx:tail}

    In this appendix we describe the methodology used to estimate the tail exponent $\alpha$ 
    of heavy-tailed distributions.  
    Naively, one might attempt to fit a straight line in log--log coordinates to the survival function
    \begin{equation}
        \mathbb{P}[X > x] \sim C x^{-\alpha},
    \end{equation}
    and take the slope as an estimate of $\alpha$.  
    However, such linear regression on doubly logarithmic plots is statistically inefficient and strongly biased:
    the choice of fitting range is arbitrary, correlations between points are ignored, 
    and finite-sample fluctuations often dominate the visual slope.  
    For these reasons, modern approaches rely instead on maximum-likelihood estimation (MLE) 
    combined with goodness-of-fit diagnostics.  
    In our work, we adopt the Clauset--Shalizi--Newman (CSN) method~\cite{Clauset2009}, 
    which has become a standard in the analysis of heavy-tailed data.
    \newline

    \noindent\textbf{Pareto Maximum-Likelihood Estimator.}

    Suppose we fix a threshold $x_{\min}$ above which the data are assumed to follow a Pareto law.  
    Given $N$ observations $\{x_i \geq x_{\min}\}$, the log-likelihood of a Pareto tail is maximized by
    \begin{equation}
        \hat{\alpha}(x_{\min}) 
        = \frac{N}{\sum_{i=1}^N \ln\!\left(\frac{x_i}{x_{\min}}\right)}.
    \end{equation}
    This estimator is unbiased in the limit $N \to \infty$, and is statistically more efficient 
    than slope-fitting in log--log space.
    \newline

    \noindent\textbf{Choosing the Threshold.}

    A central difficulty is the choice of $x_{\min}$: too low and the data deviate from a pure power law, 
    too high and too few points remain in the tail.  
    The CSN method selects $x_{\min}$ by minimizing the Kolmogorov--Smirnov (KS) distance 
    between the empirical cumulative distribution function (CDF) of the tail 
    and the fitted Pareto CDF.  
    This balances goodness-of-fit with statistical power.
    \newline

    \noindent\textbf{Goodness-of-Fit and Uncertainty.}

    To assess whether the Pareto model is plausible, the CSN method uses a bootstrap procedure:  
    synthetic datasets are generated from the fitted model and refitted in the same way, 
    yielding a distribution of KS statistics.  
    The $p$-value is the fraction of synthetic datasets with KS statistic at least as large as the empirical one.  
    This provides a rigorous test of the hypothesis ``the tail follows a power law''.  
    In addition, confidence intervals for $\alpha$ can be obtained from the bootstrap samples.
    \newline

    \noindent\textbf{Alternative Estimators.}

    Another widely used approach is the Hill estimator, which directly uses the top $k$ order statistics:  
    \begin{equation}
        \hat{\alpha}_{\mathrm{Hill}}(k) 
        = \frac{k}{\sum_{i=1}^k \ln\!\left(\frac{x_{(i)}}{x_{(k)}}\right)},
    \end{equation}
    where $x_{(i)}$ denotes the $i$-th largest observation.
    While useful for exploratory analysis, it requires careful tuning of $k$ 
    and can be sensitive to sample variability.  
    In contrast, the CSN method automates the threshold selection and provides statistical tests.
    \newline

    \noindent\textbf{Summary.}

    In summary, our estimation procedure consists of:
    \begin{enumerate}
        \item Sorting the data and considering candidate thresholds $x_{\min}$.
        \item For each threshold, estimating $\alpha$ by MLE and computing the KS statistic.
        \item Selecting the $x_{\min}$ that minimizes the KS distance.
        \item Validating the fit and estimating confidence intervals via bootstrap resampling.
    \end{enumerate}

    This methodology avoids the pitfalls of linear regression in log--log space, 
    yields statistically principled estimates of the tail exponent, 
    and allows us to assess both parameter uncertainty and goodness-of-fit.  
    In the numerical analysis, all reported values of $\alpha$ are obtained using this CSN framework.

\end{document}